\documentclass[aip,jap,preprint,groupedaddress]{revtex4-1}
\usepackage{graphicx}
\usepackage{hyperref}
\usepackage{multirow}
\newcommand*\rfrac[2]{{}^{#1}\!/_{#2}}
\begin{document}
\title{Shubnikov\,-\,de Haas oscillations, weak antilocalization effect and large linear magnetoresistance in the putative topological superconductor LuPdBi}
\author{Orest Pavlosiuk} \author{Dariusz Kaczorowski} \author{Piotr Wi{\'s}niewski*}
\affiliation{Institute for Low Temperatures and Structure Research,
Polish Academy of Sciences, P. O. Box 1410, 50-950 Wroc{\l}aw, Poland}
\date{\today}

\begin{abstract}
We present electronic transport and magnetic properties of single crystals of semimetallic half-Heusler phase LuPdBi, having theoretically predicted band inversion requisite for nontrivial topological properties. The compound exhibits superconductivity below a critical temperature $T_{\rm c}=1.8\,$K, with a zero-temperature upper critical field $B_{\rm c2}\approx2.3\,$T. Although superconducting state is clearly reflected in the electrical resistivity and magnetic susceptibility data, no corresponding anomaly can be seen in the specific heat. Temperature dependence of the electrical resistivity suggests existence of two parallel conduction channels: metallic and semiconducting, with the latter making negligible contribution at low temperatures. The magnetoresistance is huge and clearly shows a weak antilocalization effect in small magnetic fields. Above about 1.5~T, the magnetoresistance becomes linear and does not saturate in fields up to 9~T. The linear magnetoresistance is observed up to room temperature. Below 10~K, it is accompanied by Shubnikov-de Haas oscillations. Their analysis reveals charge carriers with  effective mass of $0.06\,m_e$ and a Berry phase very close to $\pi$, expected for Dirac-fermion surface states, thus corroborating topological nature of the material.
\\\\ $\,$* Correspondence to: p.wisniewski@int.pan.wroc.pl
\end{abstract}

\maketitle
Large family of half-Heusler compounds that crystallize in a noncentrosymmetric cubic MgAgAs-type structure attracts much attention due to their remarkable magnetic and electrical transport properties.\cite{Graf2011a, Casper2012a} Recent {\em ab initio} electronic structure calculations have predicted that several dozen of half-Heusler compounds, due to a strong spin-orbit coupling, have inverted bands requisite for topological properties.\cite{Lin2010,Al-Sawai2010,Chadov2010a} Topological insulators (TI) constitute a new class of materials, which are insulating in the bulk, but at the same time their surface states are protected from backscattering by Z$_2$ topology.\cite{Roy2009,Fu2007,Moore2007} These surface states are theoretically described as massless Dirac fermions with linear dispersion and lifted spin-degeneracy. Such properties open new prospects for spintronic applications. In contrast to two-dimensional (2D) TIs, like HgTe, some of half-Heusler phases possess not only band inversion at an odd number of time-reversal-invariant momenta, but also may have a bulk band-gap opening under uniaxial pressure, which lowers the crystal symmetry.\cite{Ando2013} In result, the rare-earth ($RE$) based half-Heusler compounds ($RE$PdBi, $RE$PtBi, $RE$PdSb, $RE$PtSb) may possibly form the biggest group of the three-dimensional (3D) TIs.

After the discovery of the 3D TI systems, a quest for superconductivity in materials with the non-trivial topology of electron bands started due to their potential applications in topological quantum computing.\cite{Fu2008,Tanaka2009, Akhmerov2009} Topological superconductor (TS) is a material characterized by protected Majorana surface states and bulk consisting of mixed-parity Cooper-pair states \cite{Hasan2010a,Qi2011a}. The first theoretically forecasted and experimentally discovered TS was Bi$_2$Se$_3$ doped with Cu atoms \cite{Fu2010,Sasaki2011}. Since it remains problematic to clarify the nature of the surface Majorana fermions in the inhomogeneous crystals of Cu$_x$Bi$_2$Se$_3$ (Ref.~\onlinecite{Kriener2011b}), new TS materials are sought extensively among low-carrier-density semiconductors, whose Fermi surfaces are centered around time-reversal-invariant momenta.\cite{Hsieh2012} In this context, the superconductivity recently found in the bismuthides $RE$PtBi: LaPtBi ($T_{\rm c}= 0.9\,$K),\cite{Goll2002} LuPtBi ($T_{\rm c}= 1.0\,$K), \cite{Tafti2013} and YPtBi ($T_{\rm c}= 0.77\,$K) \cite{Butch2011a,Bay2012,Bay2014} appears most attracting. Besides, the content of magnetic rare-earth may bring about other properties, such as long-range antiferromagnetic ordering or heavy fermion behavior.\cite{Canfield1991}

Potentially, the rare-earth palladium-bearing bismuthides $RE$PdBi are equally interesting from the point of view of topological and superconducting properties. Magnetic and transport behaviors of these compounds have been first investigated on polycrystalline samples. \cite{Riedemann1996,Gofryk2005,Kaczorowski2005,Gofryk2011} Most of them have been characterized as local-moment antiferromagnets (with N\'eel temperatures \mbox{$T_{\rm N}=2-13\,$K)} with electrical properties characteristic of semimetals or narrow-band semiconductors. Recently, diamagnetic YPdBi has been reported to exhibit large linear magnetoresistance (LMR) and Shubnikov-de Haas (SdH) quantum oscillations at low temperatures, and its nontrivial topological nature has been evaluated.\cite{Wang2013} In turn, CePdBi has shown some features of incipient superconductivity ($T_{\rm c}=1.3\,$K) emerging in the antiferromagnetic state ($T_{\rm N}=2\,$K), and a hypothesis on topological character of the compound has also been formulated.\cite{Goraus2013} Similarly, ErPdBi has been reported to show the coexistence of antiferromagnetism ($T_{\rm N}=1.06\,$K) and superconductivity ($T_{\rm c}=1.22\,$K), and hence acclaimed as a new platform for the research of the reciprocity of magnetic order, superconductivity, and nontrivial topological states.\cite{Pan2013a} However, in our own study on single-crystalline ErPdBi, we could not confirm those findings because no clear-cut evidence for any intrinsic superconducting state in this material has been observed.\cite{Pavlosiuk2015} Most recently, superconductivity ($T_{\rm c}=1.7\,$K) has been reported for nonmagnetic LuPdBi, accompanied by a 2D weak antilocalization (WAL) effect.\cite{Xu2014a} Since the latter feature is considered as a fingerprint of topological surface states, the compound was suggested to be a topological superconductor with Majorana edge states.\cite{Xu2014a} In this work, we present the results of our comprehensive magnetic susceptibility, specific heat and electrical resistivity measurements performed on high-quality single crystals of LuPdBi. The obtained data are here critically compared with those reported in Ref.~\onlinecite{Xu2014a}.
\section*{Results}
\subsection*{Electrical resistivity and Hall effect}
\begin{figure}
\includegraphics[height=7cm]{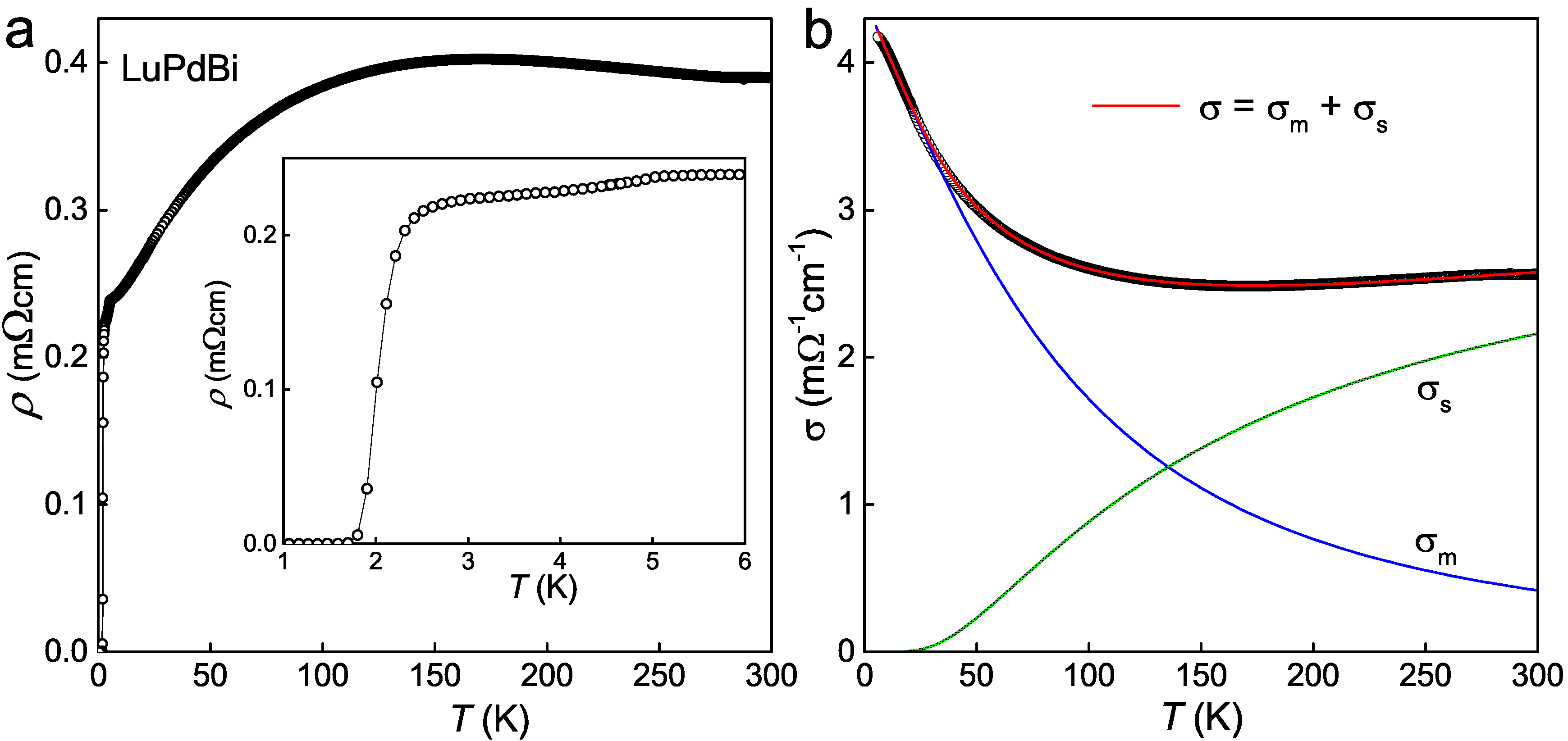}\\
\caption{{\bf Electrical transport properties of single-crystalline LuPdBi.} (a) Temperature dependence of the electrical resistivity up to 300 K. Inset:
low-temperature resistivity data revealing a superconducting transition near 2~K. (b) Temperature dependence of the electrical conductivity in the normal state
fitted by the function $\sigma (T) = \sigma _{\rm m} (T) + \sigma _{\rm s} (T)$ (red line) described in the text. The metallic and semiconducting contributions are represented by blue and green curves, respectively.
\label{RvsTplot}}
\end{figure}
The temperature dependence of the electrical resistivity, $\rho$, of the LuPdBi single crystal is shown in Figure~\ref{RvsTplot}a. From 300~K down to 
$\approx\,$170~K, the resistivity displays a semiconducting-like character (d$\rho(T)$/d$T$ $<$ 0), then at lower temperatures $\rho(T)$ becomes typical for 
metals. Such a behavior of the resistivity is common for the $RE$PdBi compounds.\cite{Gofryk2005,Kaczorowski2005,Gofryk2011,Pan2013a,Pavlosiuk2015} As 
demonstrated in Figure~\ref{RvsTplot}b, the overall behavior of the electrical resistivity in LuPdBi can be well described over an extended range of temperature 
considering two independent channels for charge transport: semiconducting- and metallic-like. The former contribution is written as $\rho _{\rm s} (T) = 
a\exp(-E_g/k_{\rm B}T)$, where $E_g$ is an energy gap between valence and conduction bands, while the latter one is expressed as $\rho _{\rm m} (T) = \rho _0 +
bT^2 +cT$, where $\rho _0$ is the residual resistivity due to scattering on structural defects, the second term represents electron-electron scattering processes,
while the third one accounts for scattering on phonons. The total conductance, $\sigma (T) \equiv 1/\rho (T)$ is the sum $\sigma (T) = \sigma _{\rm s} (T) + 
\sigma _{\rm m} (T)$, where $\sigma _{\rm s} (T) \equiv 1/\rho _{\rm s} (T)$ and $\sigma _{\rm m} (T) \equiv 1/\rho _{\rm m} (T)$. Fitting these expressions to
the experimental data of LuPdBi in the wide temperature interval from 5 K to 300 K, yielded the parameters: $a = 296\,{\rm m\Omega cm}$, $E_g$ = 11.5~meV, 
$\rho_0=226\,{\rm m\Omega cm}$, $b=0.018\,{\rm m\Omega cm/K^2}$, and $c=1.72\,{\rm m\Omega cm/K}$. The so-derived energy gap $E_g$ is very close to 15.2~meV 
reported for LuPdBi in Ref.~\onlinecite{Riedemann1996}, and very similar to the values reported for other $RE$PdBi 
phases.\cite{Gofryk2005,Kaczorowski2005,Gofryk2011, Pan2013a,Pavlosiuk2015}

It is worth noting that the temperature variation of the resistivity of LuPdBi single crystal obtained in the present work somewhat differs from that reported 
before by Xu et al.\cite{Xu2014a}, who observed a monotonous increase of the resistivity with decreasing temperature from 300 to 2~K, and attempted to analyze 
their $\rho (T)$ data above 90~K in terms of 3D variable-range hoping model. However, also in that case, the transport at low-temperatures was attributed to the 
metallic channel (though no quantitative analysis was made).

\begin{figure}
\includegraphics[height=7cm]{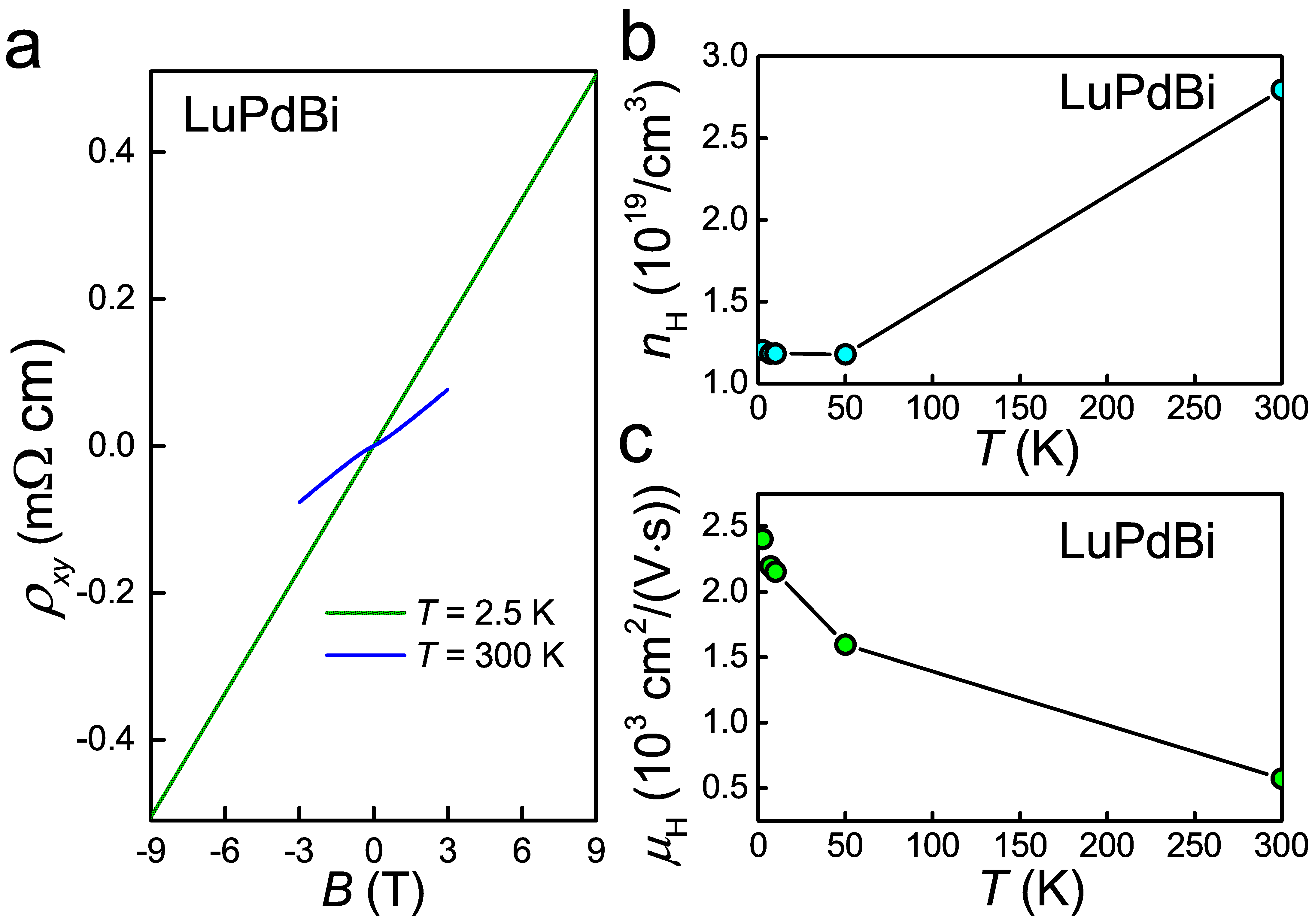}\\
\caption{\textbf{Hall effect in LuPdBi.} (a) Magnetic field dependence of the Hall resistivity measured at 2.5 and 300~K. (b) Temperature variation of the Hall
carrier concentration. (c) Temperature variation of the Hall carrier mobility.
\label{plotHall}}
\end{figure}

As displayed in Figure~\ref{plotHall}a, the Hall resistivity of LuPdBi measured at 2.5~K is a linear function of the applied magnetic field up to 9 T. The Hall 
coefficient evaluated from this data was $R_{\rm H}=0.0565\,{\rm m\Omega cm/T}$. Almost identical magnitude of $R_{\rm H}$ was derived from $\rho_{\rm xy}(B)$ 
collected at $T =$ 4, 7, 10 and 50~K (not shown). Assuming a single parabolic band, one may estimate the Hall carrier concentration, $n_{\rm H}$, to be at low 
temperatures $n_{\rm H}\simeq 1.2\times10^{19}{\rm cm^{-3}}$. Above 50 K, $n_{\rm H}$ was found to increase with increasing temperature and to attain at 300~K a 
value of $2.8\times10^{19}{\rm cm^{-3}}$ (see Figure~\ref{plotHall}b). In turn, as demonstrated in Figure~\ref{plotHall}c, the so-obtained Hall carrier mobility, $\mu_{\rm 
H}$, gradually decreases with increasing temperature, from the value $2404\,{\rm cm^2V^{-1}s^{-1}}$ at ~$T=2.5\,$K to $573\,{\rm cm^2V^{-1}s^{-1}}$ at 300~K.

The inset to Figure~\ref{RvsTplot}a displays the resistivity of single-crystalline LuPdBi at the lowest temperature studied. In agreement with 
Ref.~\onlinecite{Xu2014a}, the compound undergoes a transition to superconducting state. The resistivity starts to drop  near 2.2~K and becomes zero below 1.8~K. 
Using a criterion of 90\% drop in the resistivity value from the point where the $\rho(T)$ curve starts bending down, $T_{\rm c} = 1.9$~K was determined. This 
value is slightly higher than $T_{\rm c} =$ 1.7~K reported before for LuPdBi (Ref.~\onlinecite{Xu2014a}) and remains the highest $T_{\rm c}$ found so far for the $RET$Bi superconductors. 
As can be inferred from the inset to Figure~\ref{RvsTplot}a, besides the superconducting phase transition, we observed a small resistivity drop 
below 5~K. The existence of a fairly similar anomaly in $\rho(T)$ near this temperature has previously been reported for a few other half-Heusler bismuthides and 
antimonides,\cite{Gofryk2005,Gofryk2007} and tentatively attributed to superconductivity of thin films of metallic Bi(Sb) located at grain boundaries, yet the 
actual nature of this feature remains unclear.
\begin{figure}
\includegraphics[height=6cm]{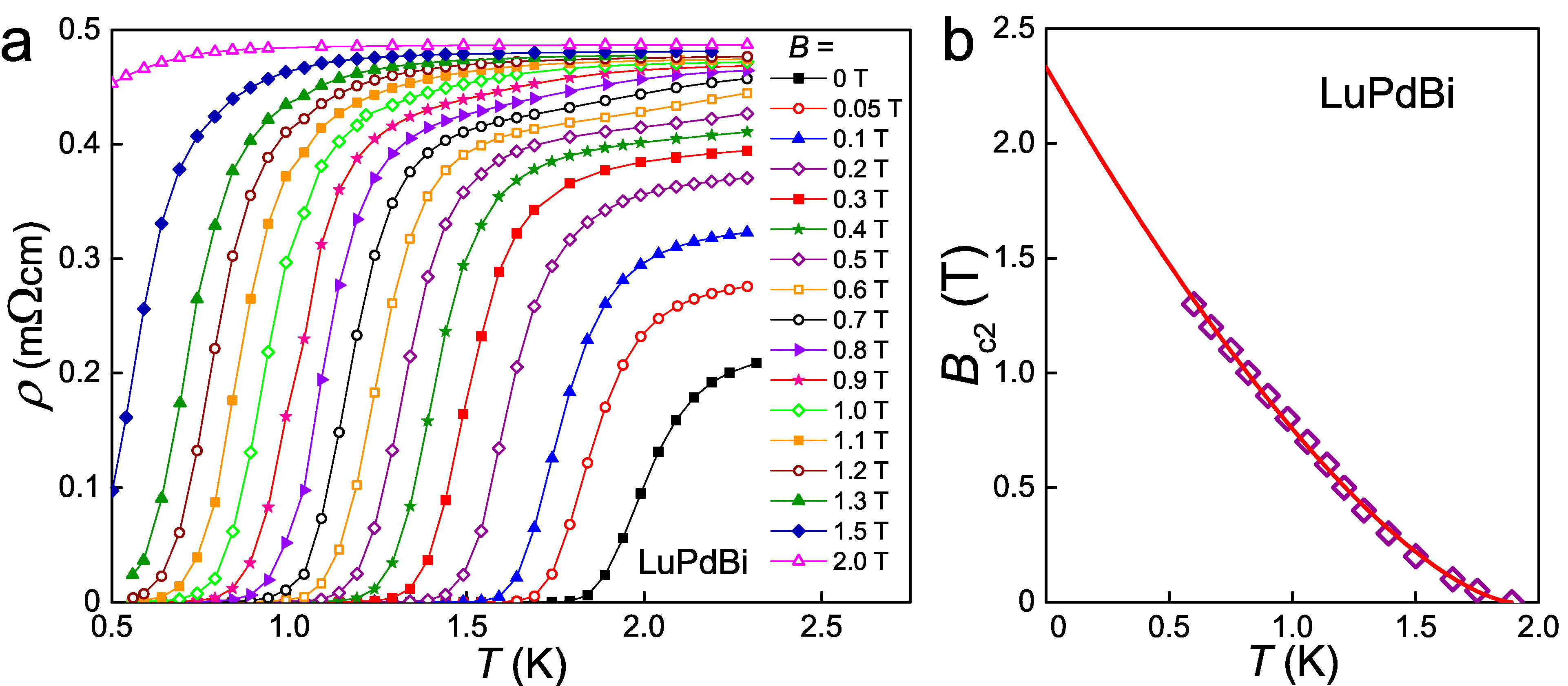}\\
\caption{\textbf{Superconductivity in LuPdBi.} (a) Low-temperature dependencies of the electrical resistivity measured in different external magnetic fields. (b) Upper critical field as a function of temperature. Solid line represents the fit of the function $B_{\rm c2}(T)=B_{\rm c2}^*(1 - T/T_{\rm c})^{1 + \alpha}$ described in the text.
\label{RvsTB}}
\end{figure}

The temperature dependence of the electrical resistivity of LuPdBi measured in external magnetic fields up to 2~T is shown in Figure~\ref{RvsTB}a. The width of 
superconducting transition increases gradually from 0.4~K at zero field to 0.7~K at 1~T. The critical temperature defined by the drop of the resistivity to 90\% 
of its normal-state value varies with magnetic field in a manner presented in Figure~\ref{RvsTB}b. The so-derived $B_{\rm c2}(T)$  dependence has a positive 
curvature, characteristic of two-band clean-limit type-II superconductors, like YNi$_2$B$_2$C, LuNi$_2$B$_2$C (Ref. \onlinecite{Freudenberger1998}) or 
MgB$_2$ (Ref.~\onlinecite{Mueller2001}). Though our sample exhibits the residual resistivity ratio $\rho$(300~K)/$\rho(4$~K) of only 1.8, the observation of SdH oscillations 
described below, indicates its very good quality, which allows us to consider the observed superconductivity to have a clean limit character. Indeed, as displayed 
in Figure~\ref{RvsTB}b, the experimental $B_{\rm c2}(T)$ data can be well approximated in the whole temperature range by the expression $B_{\rm c2}(T)=B_{\rm 
c2}^*(1 - T/T_{\rm c})^{1 + \alpha}$, previously applied, e.g., to MgB$_2$ (Ref.~\onlinecite{Mueller2001}). The fitting parameter $B_{\rm c2}^*$ = 2.3~T can be considered as
the upper limit for the upper critical field $B_{\rm c2}$(0). The parameter $\alpha$ = 0.496(22) is significantly larger than 0.32 and 0.24 (0.25) reported for 
MgB$_2$ and LuNi$_2$B$_2$C (YNi$_2$B$_2$C), respectively.\cite{Mueller2001,Freudenberger1998} Using the relation $\xi_0=\sqrt{\Phi_0/2\pi B_{\rm c2}(0)}$, one 
obtains an estimate for the coherence length $\xi_0=17$~nm. This value is about three times larger than those reported for MgB$_2$ (Ref.~\onlinecite{Mueller2001}),
YNi$_2$B$_2$C and LuNi$_2$B$_2$C (Ref.~\onlinecite{Shulga1998}).
Nevertheless, it is still distinctly smaller than the mean free path $l = 60\,$nm derived from the SdH data (see below) and this comparison corroborates our 
assumption that the superconductivity in LuPdBi occurs in the clean limit ($l/\xi_0 > 1$). The Werthamer-Helfand-Hohenberg approximation\cite{Werthamer1966} 
provides an estimate for the orbitally limited upper critical field  $B_{\rm orb}= 0.72T_{\rm c} [-{\rm d}B_{\rm c2}/{\rm d}T ]_{T_{\rm c}}$ to be 0.5 T. The 
Pauli limited upper critical field $B_{\rm P}$ can be evaluated from the equation $B_{\rm P} = \Delta/\sqrt{2}\mu_{\rm B}$, where the BCS energy gap 
$\Delta=1.76\,k_{\rm B}T_{\rm c}$. The estimate for the measured crystal of LuPdBi is $B_{\rm P}=\,$3.5~T. Thus, the Maki parameter $\alpha_{\rm M}=B_{\rm 
orb}/B_{\rm P}$ = 0.2 was obtained. One can conclude that the superconductivity in LuPdBi is Pauli limited. It is worthwhile recalling that the relationship
$B_{\rm orb} < B_{\rm c2} < B_{\rm P}$ was established also for the related half-Heusler superconductors YPtBi and LuPtBi (Refs.~\onlinecite{Bay2012, Tafti2013}).
\subsection*{Thermodynamic properties}
\begin{figure}
\includegraphics[width=14cm]{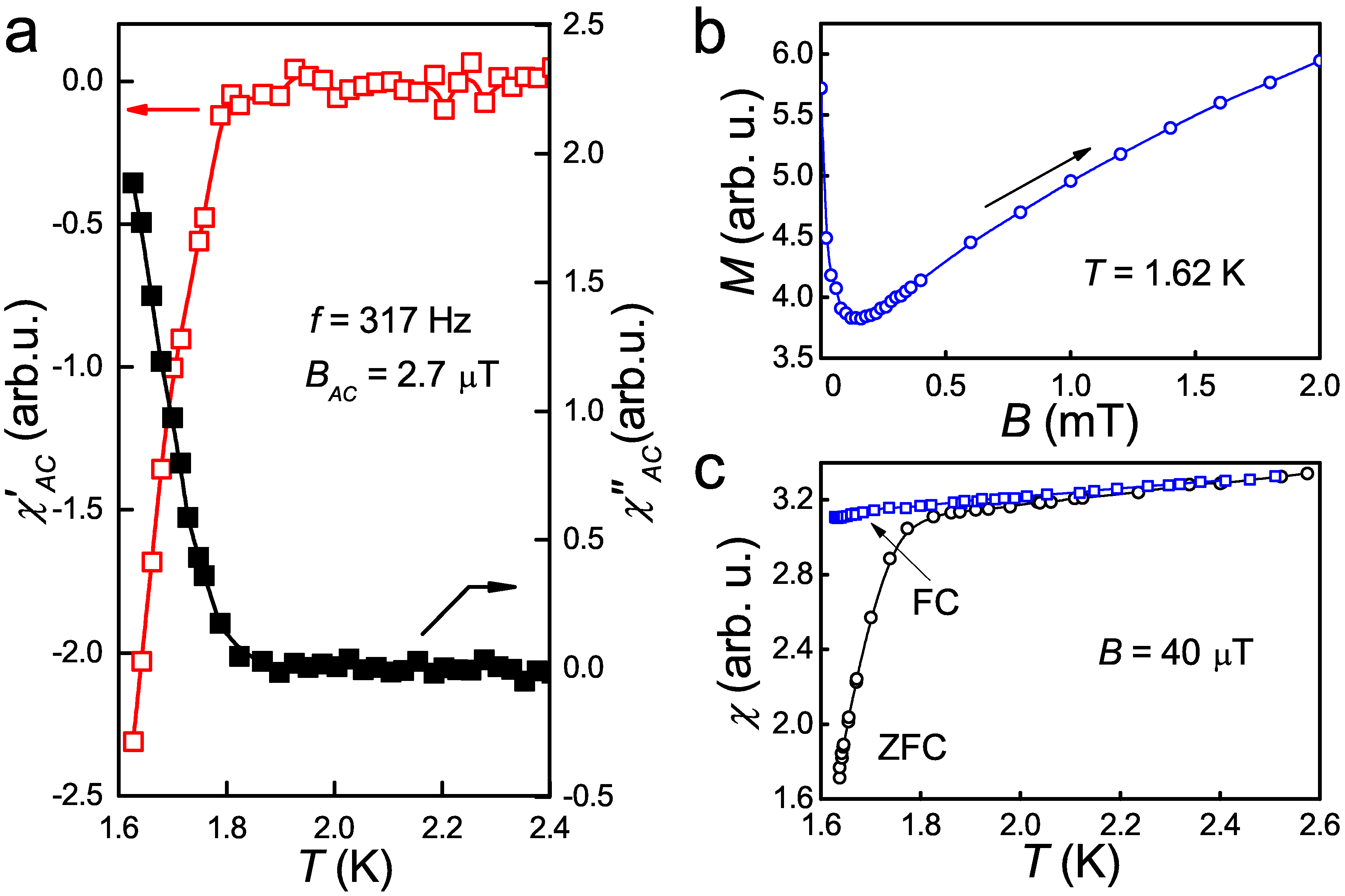}
\caption{\textbf{Magnetic properties of single-crystalline LuPdBi.} (a) Temperature variations of the real and imaginary components of the AC magnetic susceptibility taken with a magnetic field of $2.7\,\mu$T alternating with a frequency of 317~Hz. (b) Field dependence of the magnetization measured at $T=1.62$~K upon cooling the specimen in zero magnetic field.(c) Low-temperature dependence of the DC magnetic susceptibility recorded upon cooling the specimen in zero magnetic field (ZFC) and in an applied field of $40\,\mu$T (FC). All solid lines are guides to the eye.}\label{mpmsplot}
\end{figure}
Figure~\ref{mpmsplot} displays the results of magnetic measurements performed on a collection of several single crystals of LuPdBi. Both components of the AC 
magnetic susceptibility show clear anomalies at 1.8~K typical for a superconducting transition. The corresponding feature is seen on the temperature dependence of 
the DC magnetic susceptibility measured in a weak magnetic field upon cooling the sample in zero field. Comparison with the data obtained in field-cooled regime 
reveals superconducting current screening effect. Furthermore, the magnetization $M$ measured at the lowest temperature $T= $1.62~K attainable in our experimental 
setup exhibits a field variation characteristic of type-II superconductors. All these features are in concert with the electrical resistivity data and manifest 
the emergence of the superconducting state in LuPdBi below $T_{\rm c}= 1.8(1)\,$K. From the $M(B)$ curve presented in Figure~\ref{mpmsplot}b one can roughly 
estimate the upper limit for the lower critical field $B_{c1}$ to be of 0.05~mT. Then, the formula $B_{\rm c1}=\Phi_0/2\pi\lambda_{\rm sc}^2$ yields the 
penetration depth $\lambda_{\rm sc}=3629\,$nm.

\begin{figure}
\includegraphics[height=6cm]{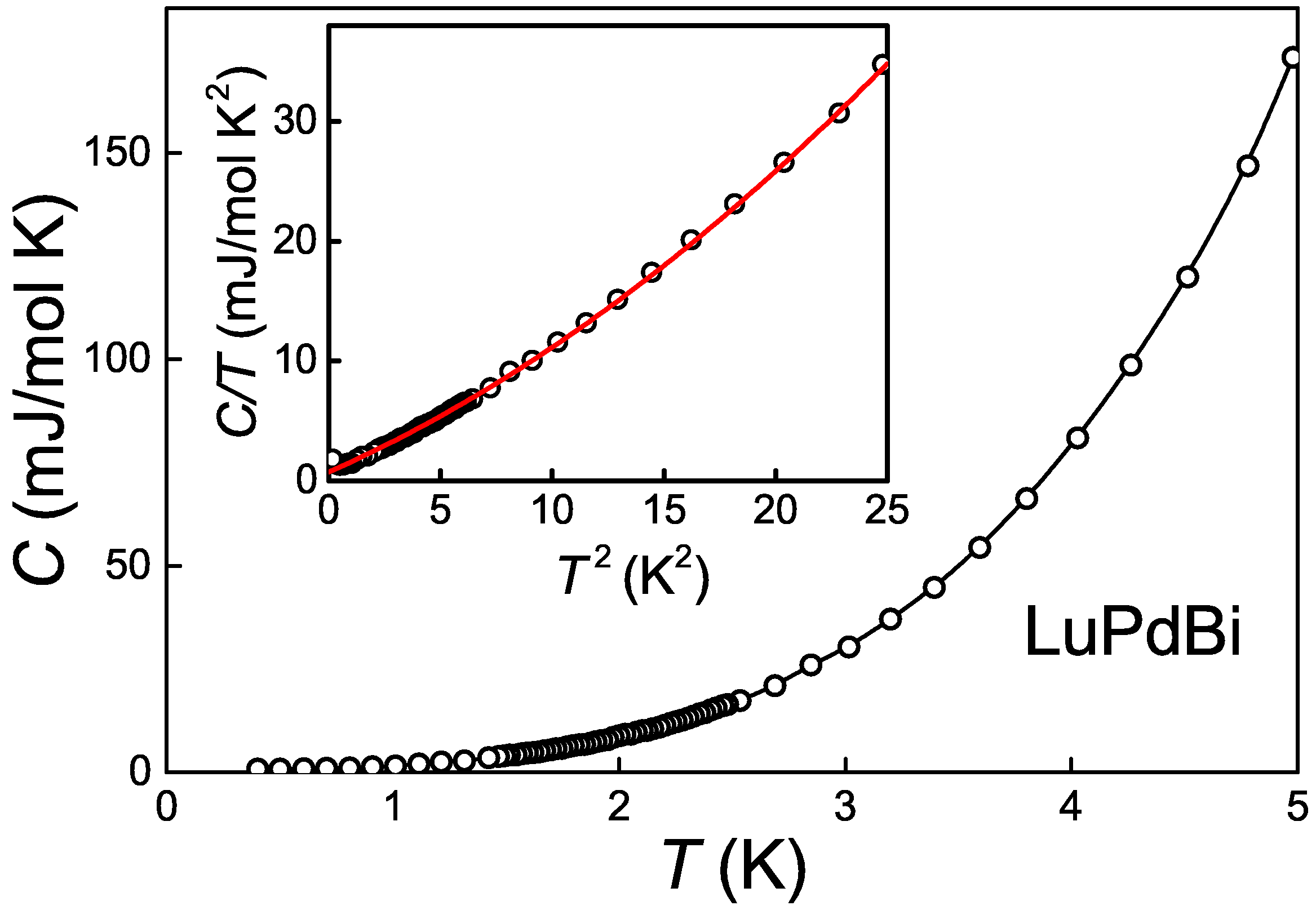}
\caption{\textbf{Heat capacity of LuPdBi.} Low-temperature dependence of the specific heat. Inset: specific heat over temperature ratio as a function of temperature squared. Solid red line represents the fit to the function $C/T = \gamma + \beta T^2 + \delta T^4$ discussed in the text. \label{HCplot}}
\end{figure}

As can be inferred from Figure~\ref{HCplot}, the low-temperature dependence of the specific heat of LuPdBi is featureless. In particular, no obvious anomaly is 
seen at the superconducting phase transition. Remarkably, analogous situation occurs for other half-Heusler superconductors: LuPtBi (Ref.~\onlinecite{Mun2013}) 
and YPtBi (Ref.~\onlinecite{Pagliuso2013}). Lack of any feature in $C(T)$ at $T_{\rm c}$ in these systems may be due to the fact that superconductivity originates from 
topologically protected surface states, which occupy a very small volume in relation to the volume of entire sample. The only exception is the result reported by 
Xu et al.\cite{Xu2014a} who observed for their single crystal of LuPdBi a prominent BCS-like specific heat jump at the onset of superconductivity. A critical 
comment on this data can be found in Supplementary Material.

As shown in the inset to Figure~\ref{HCplot}, below 5~K, the $C/T(T)$ variation can be well described by the standard formula $C/T = \gamma + \beta T^2 + \delta 
T^4$ with the electronic specific heat coefficient $\gamma=0.75\,{\rm mJ/mol\,K}^2$ and the coefficients of the phonon contribution $\beta =0.8\,{\rm
mJ/mol\,K}^4$ and $\delta =0.02\,{\rm mJ/mol\,K}^6$. From the relation $\Theta_{\rm D} = (12nR\pi^{4}/5\beta)^{1/3}$ ($R$ is the gas constant and $n$ is the
number of atoms per formula unit) one derives the Debye temperature $\Theta_{\rm D}=194\,$K. The obtained values of $\gamma$ and $\Theta_{\rm D}$ are close to 
those reported in Ref.~\onlinecite{Riedemann1996}: \mbox{$\gamma=0.096(327)\,{\rm mJ/mol\,K}^2$} and $\Theta_{\rm D}=194.9\,$K but strongly disagree with the 
results derived for LuPdBi by Xu et al.: $\gamma =11.9\,{\rm mJ/mol\,K}^2$ and $\Theta_{\rm D}=295\,$K (Ref.~\onlinecite{Xu2014a}). It is worth recalling that for other 
nonmagnetic $RET$Bi the values of $\gamma$ are also very small: 0.1 for YPtBi (Ref.~\onlinecite{Pagliuso1999}) and $0.06\,{\rm mJ/mol\,K}^2$ 
LuPtBi (Ref.~\onlinecite{Mun2013}).
\subsection*{Magnetoresistance}
\begin{figure*}
\includegraphics[width=14cm]{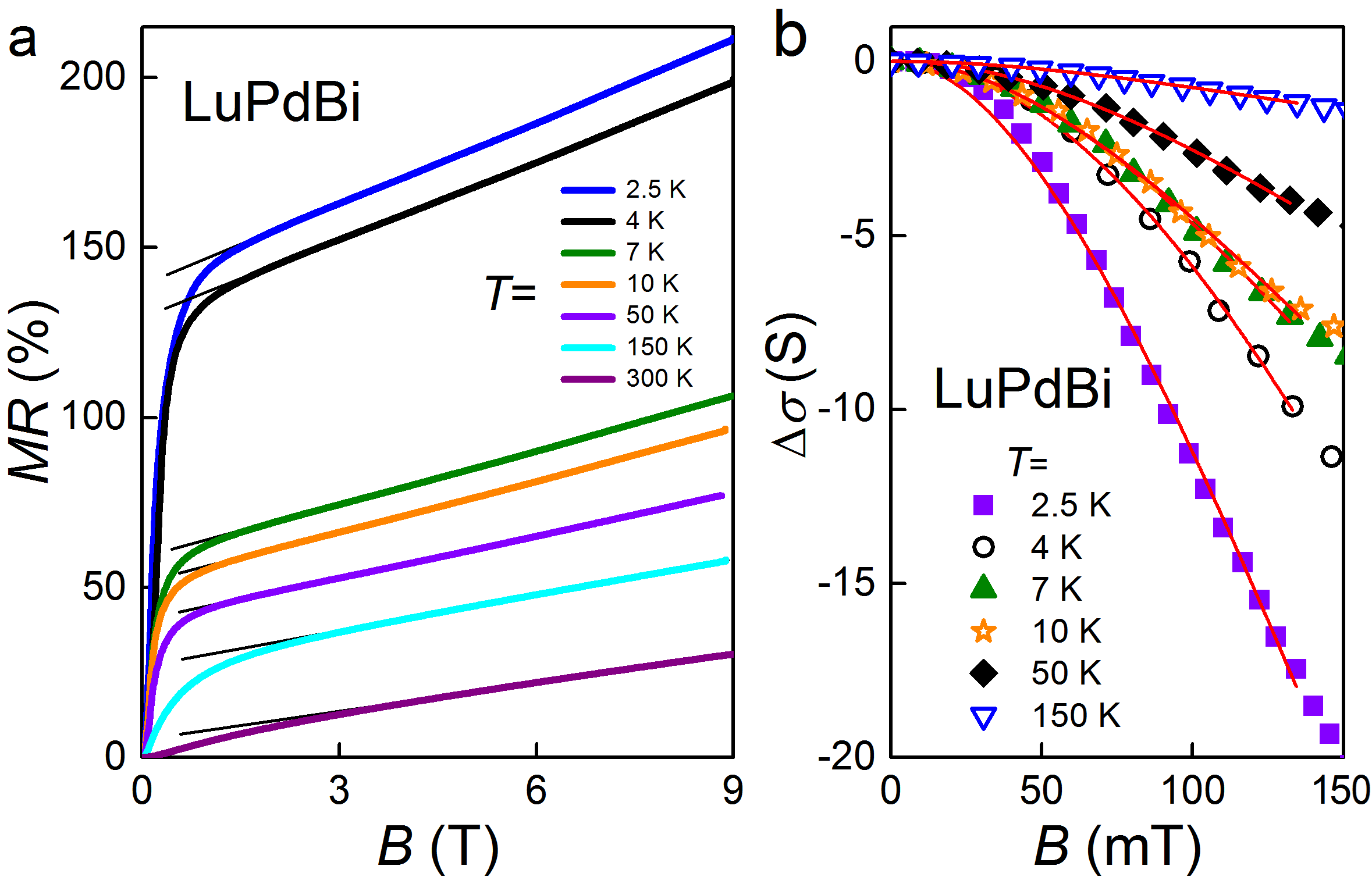}
\caption{\textbf{Transverse magnetotransport in single-crystalline LuPdBi.} (a) Field dependence of the magnetoresistance measured at different temperatures. 
Solid lines emphasize the linear behavior. (b) Low-field variations of the conductance measured at different temperatures. Solid curves represent the functions
given in Eq.~\ref{WALeq}, which account for the weak antilocalization effect. \label{MRplot}}
\end{figure*}

Figure~\ref{MRplot}a shows the magnetoresistance data of LuPdBi, defined as MR$(B)=[\rho(B)-\rho(0)]/\rho(0)]$, with $\rho(B)$ and $\rho(0)$ denoting the 
electrical resistivity measured in applied field and in zero field, respectively. Remarkably, at 2.5~K MR is as large as 210\% in the strongest field of 9~T 
achievable in our experimental setup. With increasing temperature, MR gradually diminishes, yet even at 300~K it reaches about 30\% in $B=$ 9~T, which is a very 
large value. A striking feature of MR$(B)$ is its linear nonsaturating behavior. Above a certain magnetic field (ranging from 1.5~T at 2.5~K to 3~T at 300~K), MR 
becomes proportional to $B$ with a slope systematically decreasing with increasing temperature, from about 8\%/T at the lowest temperatures, down to 3\%/T at 
300~K.

In conventional theory of metals and semiconductors, MR should grow quadratically in weak magnetic fields and then saturate in strong fields.\cite{Abrikosov1988} 
Linear magnetoresistance (LMR) is a phenomenon reported for narrow-gap semiconductors,\cite{Petrovic2003,Hu2008,Xu1997} multi-layered graphene,\cite{Friedman2010} 
topological insulators\cite{Wang2012f, Shekhar2012b,Zhang2012} and Dirac semimetals.\cite{He2014, Novak2014} Two main approaches were made to understand this 
behavior. In their classical theory, Parish and Littlewood attributed LMR to material inhomogeneities, which give rise to sturdy spatial fluctuations in the 
electrical conductivity.\cite{Parish2003} In turn, Abrikosov developed the quantum magnetoresistance theory,\cite{Abrikosov1998a} that predicts LMR in zero-gap 
band systems with linear energy dispersion being in the extreme quantum limit (EQL) in which all electrons are confined to the first Landau level. The Abrikosov 
theory was demonstrated to work correctly for graphene\cite{Friedman2010} and topological insulator Bi$_2$Se$_3$ nanosheets,\cite{Wang2012f} with topological 
surface states possessing a linear energy dispersion over wide temperature interval.

For single-crystalline LuPdBi, the detailed X-ray and microprobe characterization revealed a very good sample quality, which was further confirmed by the 
observation of SdH oscillations (see below). This prompts us to dismiss the classical theory. On the other hand, the compound shows LMR behavior virtually 
identical to that reported for silver chalcogenides, interpreted within the quantum magnetoresistance theory.\cite{Abrikosov1998a} Moreover, the SdH data 
indicated that the studied crystal was in an EQL regime at temperatures up to at least 10~K. All these findings suggest that the quantum approach is perfectly 
applicable, and the observed LMR effect is a hallmark of Dirac-like band structure in this putative topological semimetal.

In weak magnetic fields, MR of LuPdBi displays a very abrupt rise. For example, at $T = 2.5$~K, MR attains a huge value of 150\% already in a field of 1.5~T. 
Similar situation, but with lower MR increase, occurs at higher temperatures, up to 150~K. Such sharp cusps on the MR isotherms are reminiscent of a weak 
antilocalization (WAL) effect, which relies on destructive interference of the electron wave functions between two closed paths with time-reversal symmetry. In 
topologically nontrivial materials, WAL is enhanced due to ability of topologically protected surface states to collect a $\pi$ Berry phase due to the helical 
spin polarization. Thus, the WAL effect may be considered as a manifestation of surface-state transport in LuPdBi.

Taking into account the coexistence in our LuPdBi sample of two parallel conducting channels (see Figure~\ref{RvsTplot}b and the discussion above), in conjunction 
with the plausibly topological nature of the compound, one may assume that at temperatures below $\approx25$~K the conductivity was almost entirely determined by 
the metallic channel due to the sample surface. Consequently, the sheet resistance $R_s(B)$ of the conducting surface of the sample can be derived. 
Figure~\ref{MRplot}b displays the sheet magnetoconductance of LuPdBi, $\Delta\sigma(B)=(1/R_s(B))-(1/R_s(0))$, where $R_s(B)$ and $R_s(0)$ represent the data 
obtained in applied field and zero field, respectively. These experimental results can be analyzed in terms of the Hikami-Larkin-Nagaoka formula\cite{Hikami1980}
\begin{equation}
\Delta\sigma(B)=\frac{\eta e^2}{2\pi^2 \hbar}\,\Big[\psi\Big(\frac{1}{2}+\frac{\hbar}{4eL_\varphi^2B}\Big)-ln\Big(\frac{\hbar}{4eL_\varphi^2B}\Big)\Big],
\label{WALeq}
\end{equation}
where $L_\varphi$ stands for a phase coherence length, $\eta$ is a parameter depending on the strength of spin-orbit interaction and magnetic scattering, 
$\psi(x)$ is digamma function, while the other symbols have their usual meanings. For $T$ = 2.5~K, the fitted parameters are $L_\varphi=49.6(2.2)\,$nm and ${\eta e^2}/{(2\pi^2 \hbar)} 
=-153(21)$ and are comparable to $L_\varphi=95\,$nm and ${\eta e^2}/{(2\pi^2 \hbar)}\approx-10$ obtained for the same compound in Ref.~\onlinecite{Xu2014a}.  For higher temperatures, where the contribution due to the bulk channel becomes 
noticeable, the fits of Eq.~\ref{WALeq} to the experimental data are still of good quality, but the value of $\eta$ decreases with increasing temperature. More 
detailed description of the dependence of WAL parameters on temperature is given in Supplementary Material. It is worth mentioning that a similar analysis of 
MR$(B)$ of LuPdBi was presented in Ref.~\onlinecite{Xu2014a}, however the fitted values of $\eta$ were $\approx10^5$ larger than that expected for a TI system 
(see also our comment in Supplementary Material).
\subsection*{Shubnikov-de Haas oscillations}
\begin{figure}
\includegraphics[height=6.6cm]{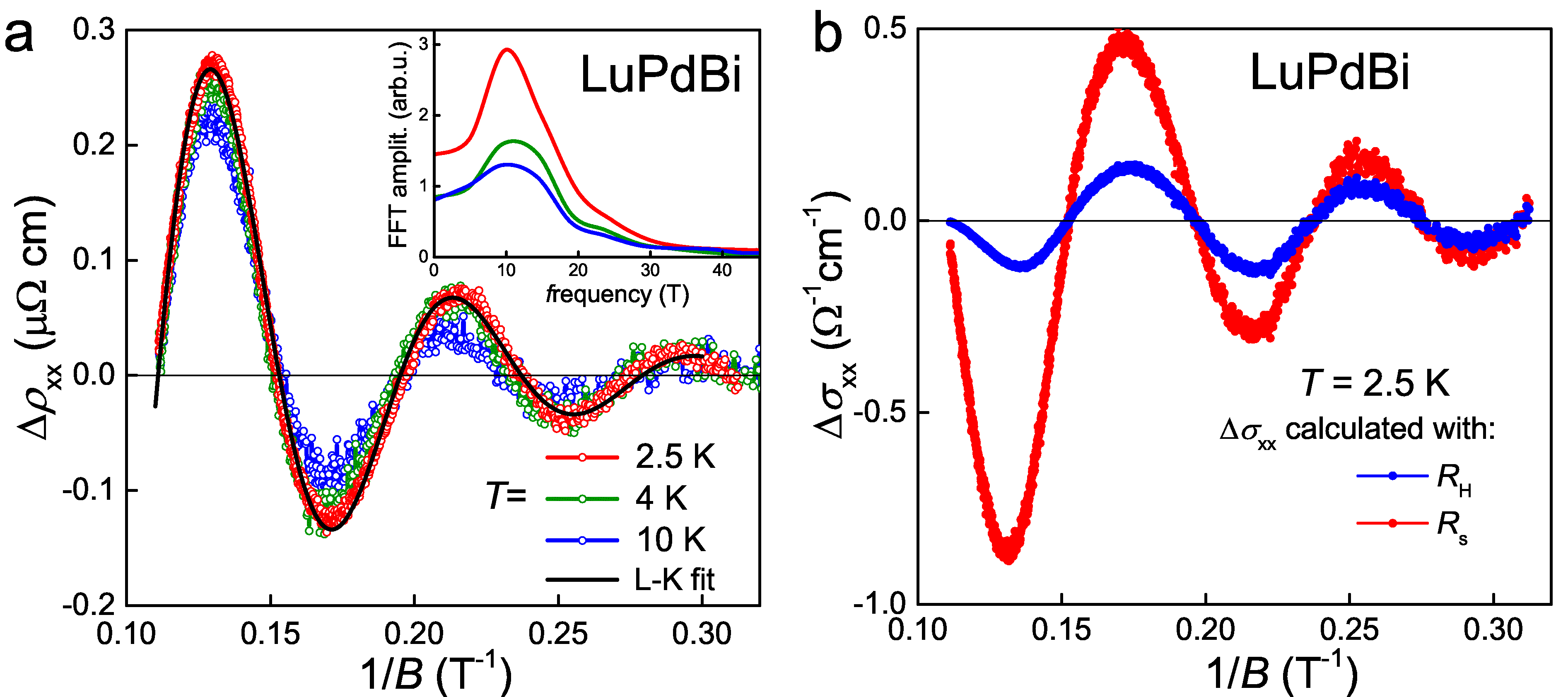}\\
\caption{{\bf Shubnikov--de Haas oscillations in LuPdBi.} (a) Oscillatory component of the electrical resistivity as a function of inverse magnetic field,
measured at different temperatures. Black solid line represents a fit of 2.5~K data with L-K formula. Inset: FFT spectra obtained for different temperatures.
(b)~Oscillatory component of longitudinal electrical conductivity at $T=2.5\,$K calculated with macroscopic Hall resistivity (blue line) or surface channel Hall
resistivity (red line). \label{SdHplot}}
\end{figure}
SdH oscillations are being recognized as a powerful tool in the studies of 3D TIs.\cite{Ando2013} Dirac fermions are characterized by a $\pi$ Berry phase (for 
ordinary metals the Berry phase is equal to zero), and the phase factors of the oscillations reveal the Berry phase values for the system. Observations of SdH 
oscillations have  been reported in some candidate topological materials from the half-Heusler family.\cite{Goll2002, Butch2011a, Wang2013, Pavlosiuk2015} For our 
LuPdBi samples, SdH oscillations were well resolved at temperatures up to 10~K, hence indicating very good quality of the single crystals measured. 
Figure~\ref{SdHplot}a shows the oscillatory component $\Delta\rho_{\rm xx}$, obtained by subtracting a background from the $\rho_{\rm xx}(B)$ isotherms. The 
oscillations are clearly periodic in 1/$B$, however their frequency is very low. Fast Fourier transform (FFT) analysis of these data yielded rather poor quality 
spectra with broad peaks corresponding to a single temperature-independent frequency $\approx 10\,$T, shown in inset to Figure~\ref{SdHplot}a. Obviously, the 
broadness of FTT peaks comes from low frequency of the oscillations and only less than three periods covered by our measurements.
In order to obtain more reliable value of the oscillation frequency $f_{\rm SdH}$ we perfomed direct fitting of the standard Lifshitz-Kosevich (L-K) expression where  $\Delta\rho_{\rm 
xx}\sim{\rm cos}[2\pi(\frac{f_{\rm SdH}}{B}-\rfrac{1}{2}+\beta)]$ with $\beta$ being a phase shift.\cite{Shoenberg1984,Ando2013} The analysis yielded $f_{\rm SdH}=11.9$~T and $\beta =0.431(2)$.

\begin{figure}
\includegraphics[height=6cm]{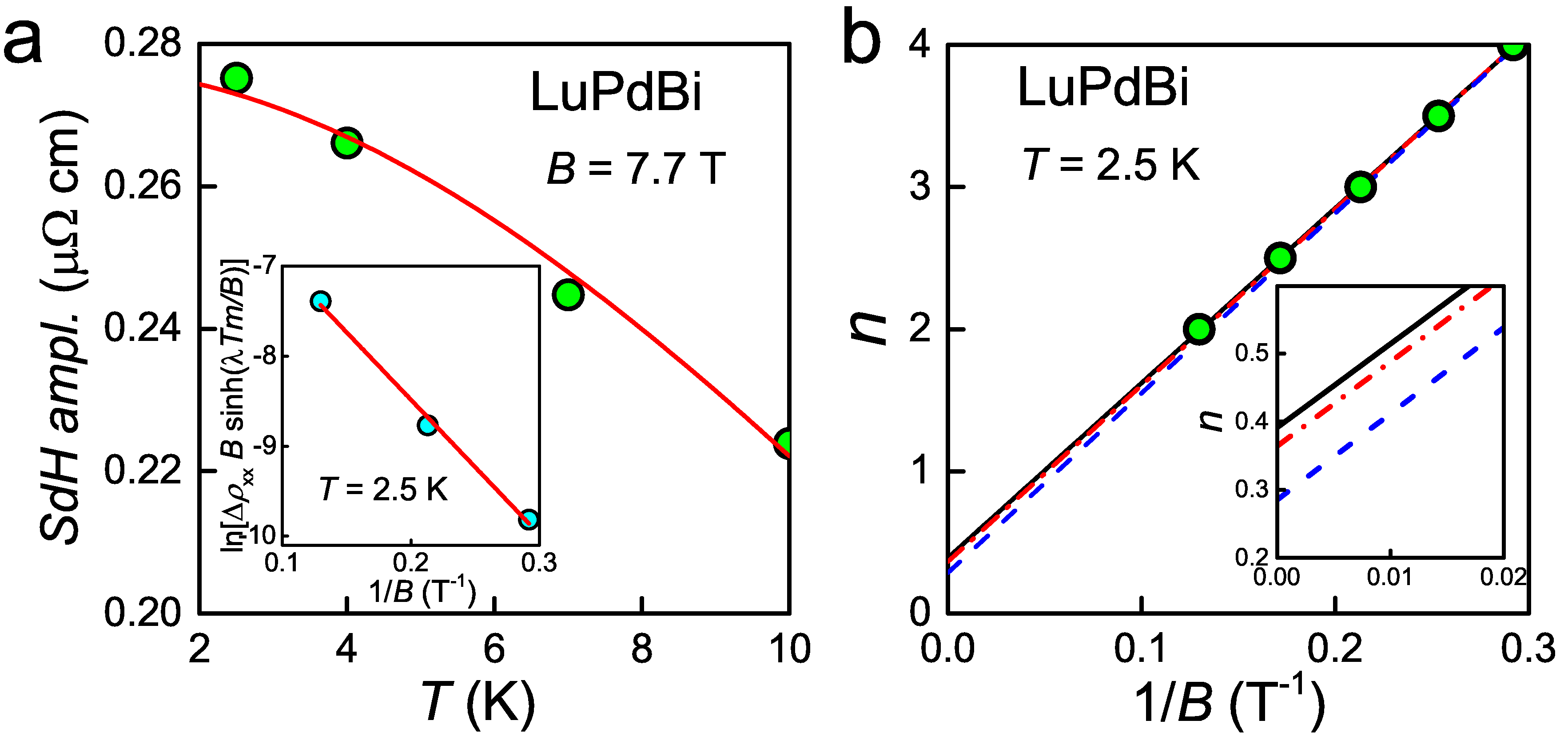}
\caption{\textbf{Analysis of the SdH oscillations in LuPdBi.} (a) Lifshitz-Kosevich description (solid line) of the temperature dependence of the SdH amplitude.
Inset: Dingle plot for the oscillations measured at 2.5~K. Solid straight line determines the Dingle temperature (see the text). (b) Landau-level fan diagram for
the SdH oscillations measured at 2.5~K. Fitted straight solid line intersecting the ordinate axis at 0.4 indicates the  Berry phase very close to $\pi$. Dashed
(blue) line represents the fit to extrema in conductivity $\Delta \sigma_{\rm xx}$ derived with macroscopic Hall coefficient, $R_{\rm H}$. Dot-dashed (red) line
corresponds to $\Delta \sigma_{\rm xx}$ derived with Hall coefficient of surface channel $R_s$ (see text for explanation).
Inset shows the blow up of the area where the fitted lines intercept ordinate axis.}
\label{massandDingle}
\end{figure}

The oscillation frequency is related to the extremal cross section area $A_{\rm F}$ of the Fermi surface  through the Onsager relation $f_{\rm SdH} = 
(\hbar/2{\pi}e)A_{\rm F}$, where $A_{\rm F}=\pi k_{\rm F}^2$. The value of $f_{\rm SdH}$ derived for LuPdBi implies the Fermi vector $k_{\rm F}=0.019\,$\AA$^{-1}$
that corresponds to a 2D carrier density $n_{\rm 2D}=k^2_{\rm F}/4\pi=2.87\times10^{11}\,$cm$^{-2}$ (we assume lifted spin degeneracy). The temperature 
dependence of the resistivity oscillations amplitude is given by the standard L-K expression $\Delta \rho_{\rm 
xx}(T)\sim\lambda(T)/\sinh(\lambda(T))$, where $\lambda(T)=2\pi^2k_{\rm B}Tm^*/\hbar eB$ and $m^*$ is the effective cyclotron mass.\cite{Shoenberg1984}
Fitting this equation to the experimental data of LuPdBi (see Figure~\ref{massandDingle}a) one obtains $m^*=0.06\,m_{\rm e}$, where $m_{\rm e}$ is the free 
electron mass. From the values of $k_{\rm F}$ and $m^*$, the Fermi velocity $v_{\rm F}= \hbar k_{\rm F}/m^* \approx\,3.6\times10^5$~m/s and the Fermi energy 
$E_{\rm F}= m^*v_{\rm F}^2 \approx\,45.7$~meV were derived.
Since calculation of the Sommerfeld coefficient for a 2D Fermi surface is impossible (it encloses zero volume), we used the formula $\gamma= k_{\rm B}^2Vm^*k_{\rm F}/3 \hbar^2$ 
(corresponding to spherical Fermi surface) and obtained the Sommerfeld coefficient equal to $0.003\,{\rm mJ/mol\,K}^2$. This value is much 
smaller than $0.75\,{\rm mJ/mol\,K}^2$ derived from the specific heat data. This finding indicates that the SdH oscillations originate from 2D states Fermi 
surface or from a tiny 3D Fermi surface, much smaller than the bulk one contributing to the specific heat. If the bulk states were contributing to SdH 
oscillations one should observe high cyclotron frequency (large $k_{\rm F}$) and/or large cyclotron mass, not seen in LuPdBi. Moreover, the Hall concentration of $n_{\rm H}\simeq 1.2\times10^{19}{\rm cm^{-3}}$ is over 50 times larger than $2.3\times10^{17}{\rm cm^{-3}}$ obtained when we assumed that SdH effect originates from 3D spherical Fermi surface. Such discrepancy in the Hall and SdH concentrations further supports the conjecture that the SdH oscillations in LuPdBi originate from the surface states.

Knowing $m^*$, one can carry out the Dingle analysis shown in the inset to Figure~\ref{massandDingle}a. The lifetime $\tau$ of the carriers can be found from the 
Dingle temperature $T_{\rm D} = \hbar/2\pi k_{\rm B}\tau$. The slope of the linear fit to the data yielded $T_{\rm D}=\,$16.9~K, which gave $\tau = 
7.2\times10^{-14}\,$s. Ergo, the mean-free path $l=v_{\rm F}\tau=26$~nm and the surface mobility $\mu_{\rm s} = e\tau/m^* = el/\hbar k_{\rm F} = 2100\, {\rm
cm}^2{\rm V}^{-1}{\rm s}^{-1}$  were obtained. Metallicity parameter $k_{\rm F}l$ of LuPdBi is then equal to 5.

Figure~\ref{massandDingle}b presents the Landau-level fan diagram for the SdH oscillations in LuPdBi at 2.5~K. Positions of the extrema in $\Delta\rho_{\rm 
xx}(1/B)$ are marked with circles as a function of their number ($n$ for maxima and $n+\rfrac{1}{2}$ for minima). Though the very small value of $f_{\rm {SdH}}$
allowed to observe only five such extrema, a good linear fit could be performed, with an intercept of 0.391(26), shown as a solid black line. However, the choice 
of resistance $\rho_{\rm xx}$ or conductance $\sigma_{\rm xx}$ for Landau level indexing may be very important and yield different phase factor.\cite{Ando2013}
Since the Hall resistivity of LuPdBi measured at 2.5~K was $\rho_{\rm xy}(B=9{\rm T})=0.54\,{\rm m\Omega cm}$, that is very close to $\rho_{\rm xx}=0.50\,{\rm m\Omega 
cm}$, it cannot be ignored in analysis of the SdH oscillations. We thus calculated $\sigma_{\rm xx}(B)=\rho_{\rm xx}(B)/(\rho_{\rm xx}^2(B)+\rho_{\rm xy}^2(B))$ and extracted the oscillatory component $\Delta\sigma_{\rm
xx}(1/B)$. These data are shown as blue points in Figure~\ref{SdHplot}b. Straight line fitted to the positions of their extrema (shown as dashed blue line in Figure~\ref{massandDingle}b) yielded the phase factor of 0.285(40) (for clarity the
datapoints are not shown).

But the case of LuPdBi is an example (which very likely can be extended to other bulk samples with topologically nontrivial states on their surface) where the 
oscillatory component of $\sigma_{\rm xx}(1/B)$ should be derived in a different manner. Our sample, as the temperature dependence of resistivity indicates, 
contains two parallel conduction channels: metallic and semiconducting (Figure~\ref{RvsTplot}b). At low temperatures, the latter makes negligible contribution to 
the conductivity, but may still play an important role in the Hall effect. However, the SdH oscillations we observe seem to originate from surface carriers. Hence, the Hall 
resistivity of the surface channel should be used for $\Delta\sigma_{\rm xx}(1/B)$ calculation. It can be easily obtained as $\rho_{s,{\rm xy}}(B)=R_sB$, where 
$R_s$ is the Hall coefficient of the surface channel, which can be derived from the formula $R_s=t/(en_{\rm 2D})$, where $n_{\rm 2D}$ is the surface carrier concentration 
and $t$ is the sample thickness.\cite{Ando2013} For our sample of LuPdBi,$R_s=0.0146\,{\rm m\Omega cm/T}$ at 2.5~K. After recalculating $\sigma_{\rm xx}(B)$ as $\rho_{\rm xx}(B)/(\rho_{\rm 
xx}^2(B)+\rho_{s,{\rm xy}}^2(B))$, extracting its oscillatory component (shown with red in Figure~\ref{SdHplot}b) and finding positions of its extrema on $1/B$
axis, we could fit another straight line resulting in the phase factor of 0.364(44) (note dot-dashed red line in Figure~\ref{massandDingle}b).

Most remarkably, all the applied methods of determinig the phase factor (we believe that the electronic nature of 3D TIs is most appropriately accounted for in that employing $R_s$)
yielded values very close to $\rfrac{1}{2}$, which was theoretically predicted for Dirac fermions. Small departure of the experimental value from $\rfrac{1}{2}$ may arise from Zeeman coupling effect\cite{Analytis2010,Chen2011} or small deviation of the energy dispersion of Dirac fermions from the linear 
one.\cite{Taskin2011b} Nevertheless, the observation in LuPdBi of a $\pi$ Berry phase is another strong indicator of the topologically nontrivial character of the compound.

\section*{Conclusions}
We synthesized high-quality single crystals and investigated the electronic properties of a putative 3D topological superconductor LuPdBi. The electrical 
transport in the compound was found to be driven by two parallel conducting channels: metallic and semiconducting, with the negligible contribution of the latter 
at low temperatures. The superconductivity emerges at $T_{\rm c} =$ 1.9~K and is characterized by the upper critical field $B_{\rm c2}(0)\leq 2.3\,$T. It is of 
type II clean-limit BCS.  Remarkably, no feature in the heat capacity is observed near the onset of superconducting state, which rather rules out its bulk 
character. Instead, we tend to attribute the superconductivity in LuPdBi to the surface states.

The compound shows the linear-in-$B$ magnetoresistance, as high as 210\% at 2.5~K and over 30\% at 300~K (in magnetic field of 9~T). In low magnetic fields, there 
clearly occurs the weak antilocalization effect. Both features are hallmarks of the topologically protected states. Furthermore, the observed SdH oscillations 
yield the very small effective mass of 0.06~$m_e$ and the Berry phase very close to $\pi$, in concert with the topological nature of LuPdBi, characterized by the 
Fermi surface sheet containing massless, extremely mobile Dirac fermions.

In summary, based on the results of our study, we postulate that at low temperatures, below about $10\,$K, LuPdBi effectively becomes a topological insulator, 
with negligible contribution of its bulk to the electronic transport and with the topological nature of its surface states being clearly reflected in the physical 
properties. The superconductivity in LuPdBi seems to emerge from the topologicaly protected surface states and further confirmation, both experimental and 
theoretical, of this conjecture is a challenging task for the future.
\section*{Methods}
\subsection*{Material preparation}
Single crystals of LuPdBi were grown from Bi flux. The elemental constituents of high purity (Lu: 99.99 wt.\%, Pd: 99.999 wt.\%, Bi: 99.9999 wt.\%), taken in 
atomic ratio 1:1:12, were placed in an alumina crucible, and sealed under argon atmosphere inside a molybdenum ampule. The reactor was heated slowly up to 
$1300\,^{\circ}$C and kept at this temperature for 15 hours. Next, it was cooled down to $1000\,^{\circ}$C at a rate of $3\,^{\circ}$C/hour, and then to room 
temperature at a rate of $5\,^{\circ}$C/hour. The crucible with the charge was extracted from the molybdenum ampule, covered with another alumina crucible filled 
with silica wool and resealed in an evacuated quartz tube. Then, the tube was heated up to 900 K at which point the excess of Bi flux was removed by 
centrifugation. The crystals obtained by this technique had a shape of cubes or plates, with dimensions up to $0.5\times0.5\times0.5\,{\rm mm^3}$ or 
$0.1\times0.5\times3\,{\rm mm^3}$, respectively. They had metallic luster and were stable against air and moisture.

\subsection*{Material characterization}
Powdered single crystals of LuPdBi were characterized at room temperature by powder X-ray diffraction (PXRD) carried out using an X'pert Pro PANanalitical 
diffractometer with Cu-K$\alpha$ radiation. The crystal structure refinements and the theoretical PXRD pattern calculations were done employing the FULLPROF 
program.\cite{Rodriguez-Carvajal1993} Quality of the samples studied was also verified by single crystal X-ray diffraction (SCXRD) performed at room temperature 
on an Oxford Diffraction Xcalibur four-circle diffractometer equipped with a CCD camera and using Mo-K$\alpha$ radiation. The SCXRD data analysis was done using 
the program package SHELXL-97.\cite{Sheldrick2008} The results of PXRD and SCXRD investigations are presented in Supplementary Material.

Chemical composition of the LuPdBi crystals was examined on a FEI scanning electron microscope (SEM) equipped with an EDAX Genesis XM4 spectrometer. The specimens 
were glued to a SEM stub using carbon tape. The crystals were found homogeneous and free of foreign phases, with the chemical compositions very close to the ideal 
one (for details see Supplementary Material).
\subsection*{Physical measurements}
Electrical resistivity and Hall measurements were carried out from 0.4 K to 300 K in applied magnetic fields up to 9~T using a conventional ac four-point 
technique implemented in a Quantum Design PPMS platform.
Dimensions of samples were: $0.010\times0.051\times0.227\,{\rm cm^3}$, for $\rho_{xx}$, and $0.0067\times0.051\times0.1524\,{\rm cm^3}$ for $\rho_{xy}$ 
measurements.
Current and voltage leads were $50\,\mu m$ thick silver wires attached to the parallelepiped-shaped specimens with silver paste and additionally spot welded. The 
heat capacity was measured on 10.6~mg collection of single crystals, in the temperature interval 0.4---5 K by relaxation time method using also the PPMS platform. 
DC magnetization and AC magnetic susceptibility measurements were performed on 30.1~mg collection of single crystals, in the temperature range 1.62---2.6 K in 
weak magnetic fields up to 2~mT employing a Quantum Design MPMS-XL SQUID magnetometer.
%
%

%
\section*{Acknowledgements}
Authors thank Dr. Marek Daszkiewicz for the X-ray characterization of single crystalline LuPdBi. The work was financially supported by the National Centre of 
Science (Poland) under research grant 2011/01/B/ST3/04466.
\section*{Author contributions}
D.K. and P.W. designed the research project, O.P. prepared the samples, O.P. and P.W. performed the physical measurements and analyzed the data. All the authors 
discussed the experimental results and wrote the manuscript.
\newcommand{\fSref}[1]{Fig.~S\ref{#1}}
\bibliographystyle{naturemag}
%
%
%
\noindent \begin{large}{\bf SUPPLEMENTARY MATERIAL} \end{large}
%
\section*{Material characterization}
\subsection*{X-ray diffraction}
The PXRD results (see Fig. S1) confirmed a single-phase character of the obtained single crystals of LuPdBi. The X-ray diffraction pattern can be fully indexed within the $F$\={4}3$m$ space group, characteristic of half-Heusler compounds, and yields the cubic lattice parameter $a = 6.565(1)\,$\AA. This value is in perfect accord with the literature value $6.566(1)\,$\AA~determined for polycrystalline sample,\cite{Haase2002} and the value $6.56\,$\AA~reported for thin films of LuPdBi.\cite{Shan2013} It is however markedly different from the lattice parameter of $6.63\,$\AA~stated by Xu et al. for their powdered single crystals.\cite{Xu2014a} As can be inferred from Fig. S1, the experimental PXRD pattern of our single-crystalline LuPdBi can be very well modeled with the MgAgAs-type crystal structure with the Lu atoms located at the crystallographic 4$a$ ($0, 0, 0$) sites, the Pd atoms occupying the 4$c$ ($1/4, 1/4, 1/4$) sites, and the Bi atoms placed at the 4$b$ ($1/2, 1/2, 1/2$) sites. The same structural model was recently shown to account well for the PXRD data of powdered single crystals of YPdBi and YPtBi.\cite{Nowak2014}
\begin{figure}[h]
\includegraphics[height=8cm]{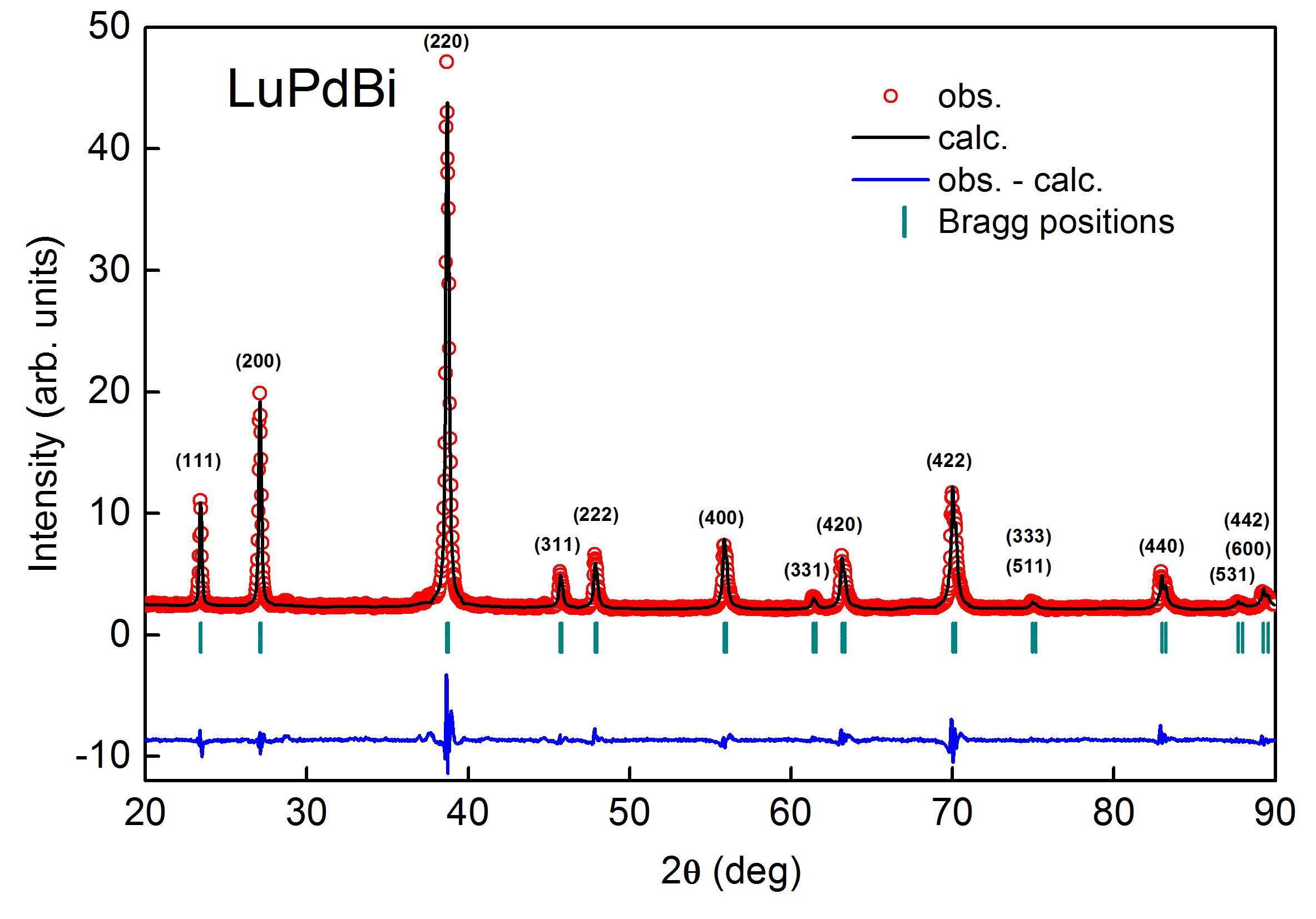}\\
\textbf{Figure S1. XRD pattern for powdered single crystals of LuPdBi.}
\label{XRDplot}
\end{figure}
\newpage 
\noindent\textbf{Table S1.}  Crystal data  and structure refinement for LuPdBi.\\
\includegraphics[width=12.5cm]{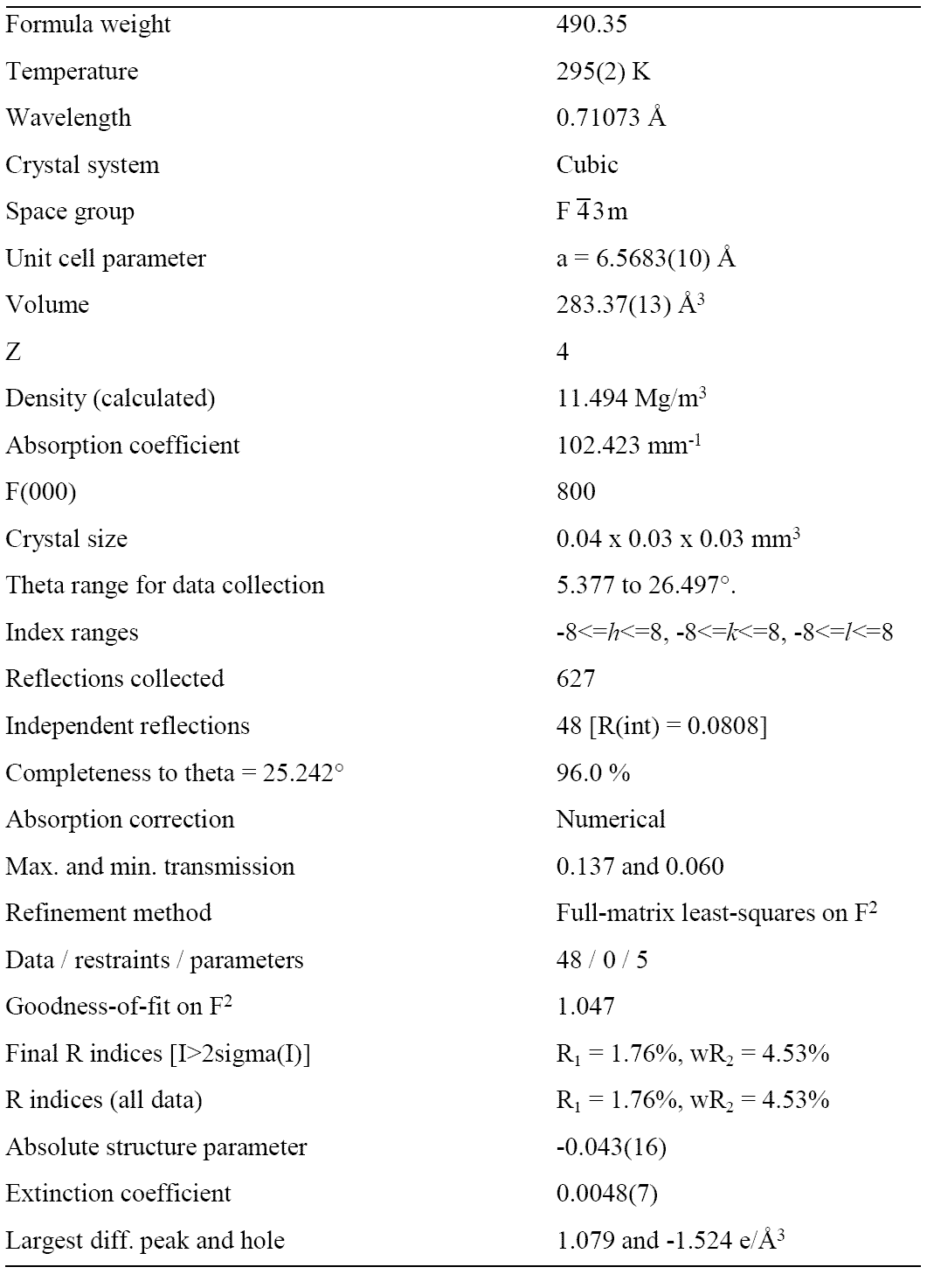}\\
\textbf{Table S2.}  Atomic coordinates and equivalent isotropic displacement parameters $U(eq)$ for LuPdBi. $U(eq)$ is defined as one third of the trace of the orthogonalized $U^{ij}$ tensor.\\
\includegraphics[width=7.4cm]{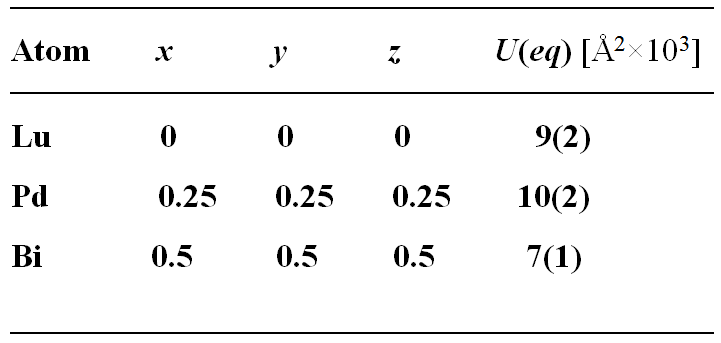}\\
It is worth noting that despite the aforementioned divergence in the values of the lattice parameter $a$, the X-ray diffractogram of LuPdBi presented in Ref.~\onlinecite{Xu2014a} appears very similar to our result displayed in Fig. S1. Small differences in the relative intensities of a few Bragg peaks [e.g., (311), (222), (420)] might be attributed to a certain level of atomic disorder in the sample studied by Xu at al., since our PXRD data were refined to a small value of the residual Bragg factor $R$ = 4.61\%, with full occupancies of all the crystallographic positions and no structural disorder (see also the discussion presented in Ref.~\onlinecite{Nowak2014}).

The crystal structure of the studied samples of LuPdBi was also checked by the single crystal X-ray diffraction method. Details on the performed refinement are gathered in Table S1. The positional data and the equivalent isotropic displacement parameters are given in Table S2. The so-refined lattice parameter $a= 6.5683(10)$~\AA is very close to that derived from the PXRD data, and thus significantly smaller than the result reported for single-crystalline LuPdBi by Xu at al.\cite{Xu2014a}

\subsection*{Energy-dispersive X-ray spectroscopy}
Examples of the EDX spectrum and the SEM image obtained for the single crystals of LuPdBi is presented in Fig. S2. The chemical composition derived from these data is Lu$_{32.00(32)}$Pd$_{33.93(33)}$Bi$_{34.07(34)}$, in a very good accord with the ideal equiatomic one.
\begin{figure}[h]
\includegraphics[height=7cm]{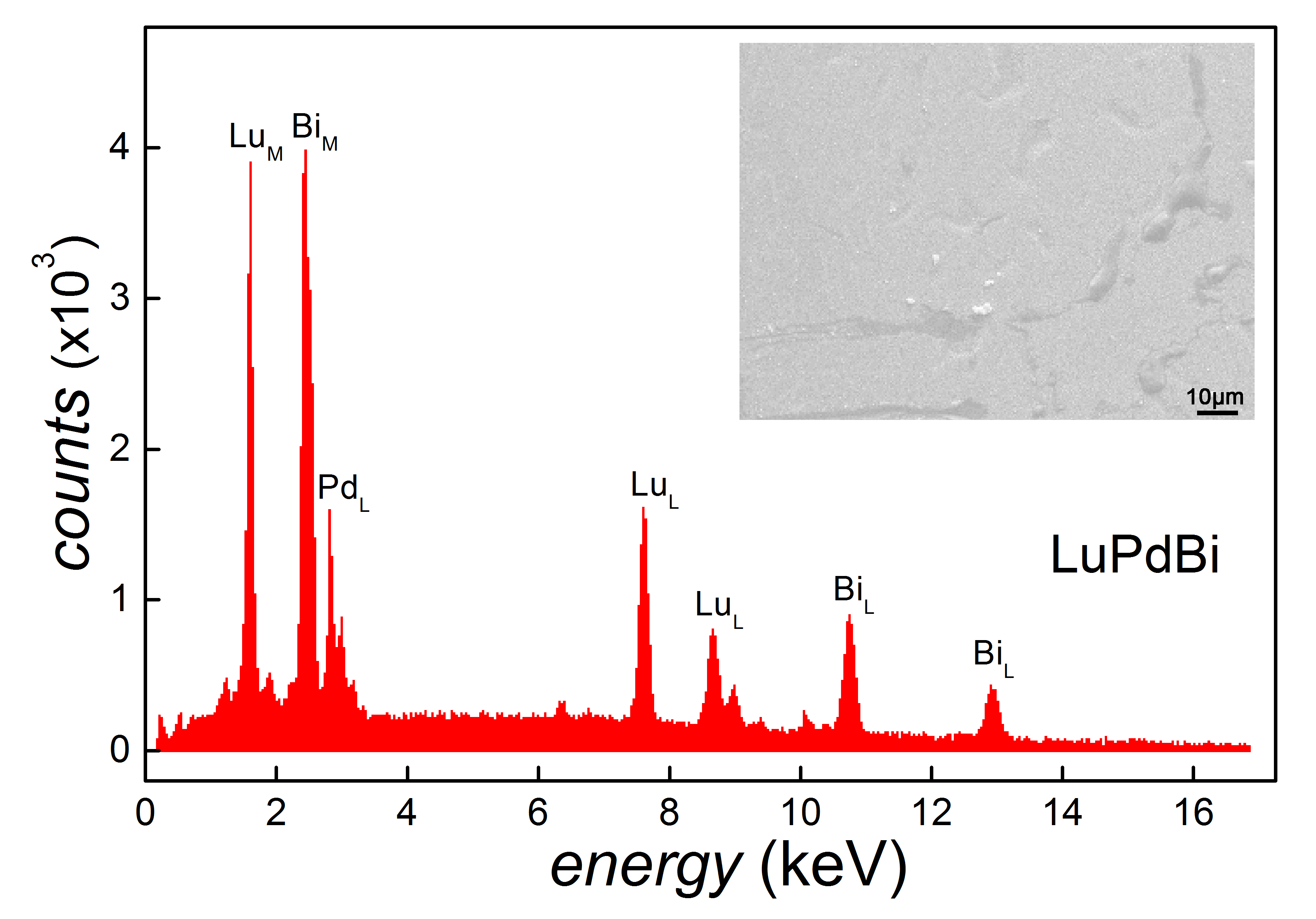}\\
\textbf{Figure S2. SEM-EDX results for single-crystalline LuPdBi.}
 \label{edx}
\end{figure}
\section*{Electronic and thermodynamic properties}
\subsection*{WAL parameters } 
The temperature dependencies of the weak antilocalization parameters $\frac{\eta e^2}{2\pi^2 \hbar}$ and $L_\varphi$ (together with their uncertainties $d(\frac{\eta e^2}{2\pi^2 \hbar})$ and $dL_\varphi$) obtained from fitting the Hikami-Larkin-Nagaoka (HLN) formula (Eq.~1 in the main text) to the experimental $\sigma(B)$ data of LuPdBi are collected in Table S3. 
When both parameters were free to vary in fitting procedure, the values given in second and fourth columns of the Table S3 were obtained. However  the $L_\varphi$ varied with temperature in irregular way and both parameters were strongly dependent on each other, as shown by statistical dependency values in fifth column. Thus we could not consider all these values of $\frac{\eta e^2}{2\pi^2 \hbar}$ and $L_\varphi$ as reliable.
Reasoning that the WAL model is most adequate at lowest temperature, where the contribution from the bulk semiconducting channel is the smallest, we repeated the fitting with the value of $\frac{\eta e^2}{2\pi^2 \hbar}$ fixed at the value derived at $T=2.5\,$K.
Resulting values of $L_\varphi$ are collected in sixth column of the Table. It is worth noting that their errors are now smaller by about one order of magnitude. These values decrease monotonously with increasing temperature, which seems to reflect the increasing of the bulk channel contribution.
\\\\
\noindent\textbf{TABLE S3.} WAL parameters for sheet conductance of LuPdBi at different temperatures.\\
\setlength{\tabcolsep}{8pt}
\renewcommand{\arraystretch}{1.2}
\begin{tabular} {l| rrrrc|cc}
\multirow{2}{*}{$T$(K)} & \multirow{2}{*}{$\frac{\eta e^2}{2\pi^2 \hbar}$} & \multirow{2}{*}{$d(\frac{\eta e^2}{2\pi^2 \hbar})$}  & \multirow{2}{*}{$L_\varphi$} & \multirow{2}{*}{$dL_\varphi$} & dependency
& \multirow{2}{*}{$L_\varphi|_{\eta(T=2.5K)}$} & \multirow{2}{*}{$dL_\varphi|_{\eta(T=2.5K)}$}\\
&&&&&  $(\frac{\eta e^2}{2\pi^2 \hbar},L\varphi)$ &&\\\hline
2.5	& -153.7 & 21.8 & 49.6 & 2.2 &	0.9976 & 49.6 & 0.1 \\
4	& -340.5 & 149.1 & 33.1 & 4.0 &	0.9996 & 41.3 & 0.1 \\
7	& -99.8  & 21.6 & 43.5 & 2.8 &	0.9986 & 38.4 & 0.1 \\
10	& -65.6  & 12.9 & 48.8 & 3.0 &	0.9977 & 37.9 & 0.1 \\
50	& -49.6  & 7.4  & 44.7 & 2.0 &	0.9984 & 32.6 & 0.1 \\
150	& -8.8   & 0.5   & 52.1 & 0.9 &	0.9971 & 23.6 & 0.1 \\
\end{tabular}
\newpage
For more clarity we plotted the values of $\frac{\eta e^2}{2\pi^2 \hbar}$ and $L_\varphi$ resulting from both fitting procedures, versus temperature, as shown in Fig. S3.
\\
\begin{figure}[t]
\includegraphics[height=8cm]{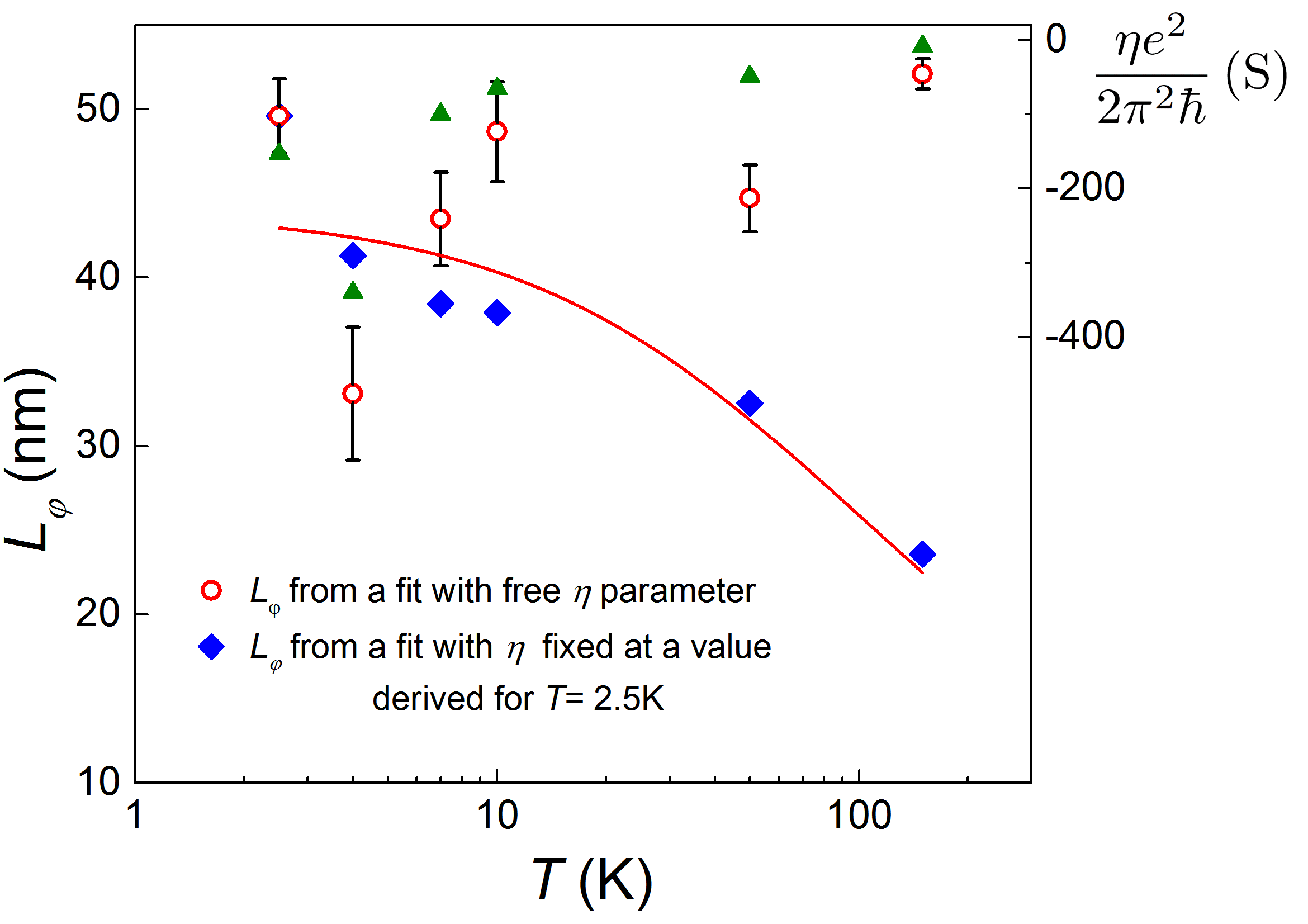}\\
\textbf{Figure S3.} Temperature variations of the WAL parameters for single-crystalline LuPdBi. Green triangles represent values of $\frac{\eta e^2}{2\pi^2 \hbar}$,  circles and diamonds correspond to $L_\varphi$ obtained from two fitting procedures described in text.
\label{WALp}
\end{figure}
The temperature variation of the phase coherence length $L\varphi|_{\eta(T=2.5K)}$ (resulting from the HLN fit with $\frac{\eta e^2}{2\pi^2 \hbar}$ fixed at the value derived at $T=2.5\,$K.) can be roughly described by the formula given in Ref.~\onlinecite{Xu2014a} (represented in Fig. S3 by the solid red curve), however such a fit yields the electron-phonon interaction parameter close to zero.

The $\frac{\eta e^2}{2\pi^2 \hbar}$ parameter derived in Ref.~\onlinecite{Xu2014a} was  about an order of magnitude smaller than that obtained on our sample, whereas corresponding values of $L_\varphi$ were about twice as large as our $L\varphi|_{\eta(T=2.5K)}$ data. This is rather a good agreement, taking into account different electronic character of the samples examined in our work and that described in Ref.~\onlinecite{Xu2014a}. 
It seems to be a common feature of half-Heusler compounds that the values of parameter $-\eta$ are of order $10^5-10^6$, thus much larger than 1/2 expected for 2D TI systems. Apart from LuPdBi it has very recently been observed in samples of LuPtSb,\cite{Hou2015} YPtBi and LuPtBi.\cite{Shekhar2015} Authors of all these papers attributed this fairly unexpected result to contributions from bulk or side wall conductivity channels. 
\subsection*{Hall parameters}
In concert with the previous report,\cite{Xu2014a} the Hall carrier concentration in the single crystal investigated in our work was found temperature independent below 50 K and then to increase with increasing temperatures. However, in the entire temperature range covered, the magnitude of $n_{\rm H}$ determined in our study is approximately four times smaller than that estimated by Xu et al. (their result was $n_{\rm H}\simeq 4.8\times10^{19}{\rm cm^{-3}}$ at low temperatures). Remarkably, our findings are completely different as regards the Hall carrier mobility. For our sample, $\mu_{\rm H}$ was found to gradually decrease with increasing temperature from the value $2404\,{\rm cm^2V^{-1}s^{-1}}$ at ~$T=2.5\,$K to $573\,{\rm cm^2V^{-1}s^{-1}}$ at 300~K, while Xu et al. reported an increase from 330 to $380\,{\rm cm^2V^{-1}s^{-1}}$ in the same temperature interval.

\subsection*{Heat capacity}
The striking result reported in Ref.~\onlinecite{Xu2014a} is the occurrence of a pronounced $\lambda$-shaped anomaly in the specific heat of LuPdBi that coincides with the superconducting transition. Consequently, Xu et al. ascribed the superconductivity in their single crystals to the bulk. This result markedly differs from the behavior of our single crystals, where no visible feature in $C(T)$ was found near $T_{\rm c}$, despite the formation of superconducting state was convincingly proved by means of the magnetic and electrical transport measurements. This finding led us to the conclusion that the superconductivity in our samples of LuPdBi is confined to the surface, with the superconducting condensate volume negligibly small as compared to the bulk. The result obtained by Xu et al. differs also from the properties of closely related systems LuPtBi and YPtBi,\cite{Mun2013,Pagliuso2013} for which the specific heat was found featureless at the superconducting transition. For the latter compound we recently confirmed the lack of any anomaly at $T_{\rm c}$ in our own investigation carried out on high-quality single crystals.\cite{Pavlosiuk_YPtBi}

The observed discrepancy might be related to the different electronic character of the single crystals studied in our work and those investigated in Ref.~\onlinecite{Xu2014a}. Comparison of the temperature-dependent electrical resistivity suggests that our samples are much more semimetallic than the predominantly semiconducting one reported by Xu et al. In this context, however, the observed over-an-order-of-magnitude discrepancy between the Sommerfeld coefficient $\gamma = 0.75{\rm mJ/mol\,K}^2$ derived in the present work and $\gamma = 11.9{\rm mJ/mol\,K}^2$ found by Xu et al. becomes most surprising. Here, again, one should note that the electronic specific heat of the related semimetallic nonmagnetic bismuthides, like LuPtBi and YPtBi (Refs.~\onlinecite{Riedemann1996, Mun2013, Pagliuso2013, Pavlosiuk_YPtBi}), and antimonides, like LuPtSb, YPdSb and YPtSb (Ref.~\onlinecite{Nowak2013}) is similar to that established by us for LuPdBi.


\begin{thebibliography}{10}
\expandafter\ifx\csname url\endcsname\relax
  \def\url#1{\texttt{#1}}\fi
\expandafter\ifx\csname urlprefix\endcsname\relax\def\urlprefix{URL }\fi
\providecommand{\bibinfo}[2]{#2}
\providecommand{\eprint}[2][]{\url{#2}}

\bibitem{Graf2011a}
\bibinfo{author}{Graf, T.}, \bibinfo{author}{Felser, C.} \&
  \bibinfo{author}{Parkin, S.~S.}
\newblock \bibinfo{title}{{Simple rules for the understanding of Heusler
  compounds}}.
\newblock \emph{\bibinfo{journal}{Prog. Solid State Chem.}}
  \textbf{\bibinfo{volume}{39}}, \bibinfo{pages}{1--50};
\newblock
  \emph{DOI:10.1016/j.progsolidstchem.2011.02.001} (\bibinfo{year}{2011}).

\bibitem{Casper2012a}
\bibinfo{author}{Casper, F.}, \bibinfo{author}{Graf, T.},
  \bibinfo{author}{Chadov, S.}, \bibinfo{author}{Balke, B.} \&
  \bibinfo{author}{Felser, C.}
\newblock \bibinfo{title}{{Half-Heusler compounds: novel materials for energy
  and spintronic applications}}.
\newblock \emph{\bibinfo{journal}{Semicond. Sci. Technol.}}
  \textbf{\bibinfo{volume}{27}}, \bibinfo{pages}{063001};
\newblock \emph{DOI:10.1088/0268-1242/27/6/063001} (\bibinfo{year}{2012}).

\bibitem{Lin2010}
\bibinfo{author}{Lin, H.} \emph{et~al.}
\newblock \bibinfo{title}{{Half-Heusler ternary compounds as new
  multifunctional experimental platforms for topological quantum phenomena.}}
\newblock \emph{\bibinfo{journal}{Nat. Mater.}} \textbf{\bibinfo{volume}{9}},
  \bibinfo{pages}{546--9};
\newblock \emph{DOI:10.1038/nmat2771} (\bibinfo{year}{2010}).

\bibitem{Al-Sawai2010}
\bibinfo{author}{Al-Sawai, W.} \emph{et~al.}
\newblock \bibinfo{title}{{Topological electronic structure in half-Heusler
  topological insulators}}.
\newblock \emph{\bibinfo{journal}{Phys. Rev. B}} \textbf{\bibinfo{volume}{82}},
  \bibinfo{pages}{125208};
\newblock \emph{DOI:10.1103/PhysRevB.82.125208} (\bibinfo{year}{2010}).

\bibitem{Chadov2010a}
\bibinfo{author}{Chadov, S.} \emph{et~al.}
\newblock \bibinfo{title}{{Tunable multifunctional topological insulators in
  ternary Heusler compounds.}}
\newblock \emph{\bibinfo{journal}{Nat. Mater.}} \textbf{\bibinfo{volume}{9}},
  \bibinfo{pages}{541--5};
\newblock \emph{DOI:10.1038/nmat2770} (\bibinfo{year}{2010}).

\bibitem{Roy2009}
\bibinfo{author}{Roy, R.}
\newblock \bibinfo{title}{{Topological phases and the quantum spin Hall effect
  in three dimensions}}.
\newblock \emph{\bibinfo{journal}{Phys. Rev. B}} \textbf{\bibinfo{volume}{79}},
  \bibinfo{pages}{195322};
\newblock \emph{DOI:10.1103/PhysRevB.79.195322} (\bibinfo{year}{2009}).

\bibitem{Fu2007}
\bibinfo{author}{Fu, L.}, \bibinfo{author}{Kane, C.~L.} \&
  \bibinfo{author}{Mele, E.~J.}
\newblock \bibinfo{title}{{Topological insulators in three dimensions.}}
\newblock \emph{\bibinfo{journal}{Phys. Rev. Lett.}}
  \textbf{\bibinfo{volume}{98}}, \bibinfo{pages}{106803};
\newblock \emph{DOI:10.1103/PhysRevLett.98.106803} (\bibinfo{year}{2007}).

\bibitem{Moore2007}
\bibinfo{author}{Moore, J.} \& \bibinfo{author}{Balents, L.}
\newblock \bibinfo{title}{{Topological invariants of time-reversal-invariant
  band structures}}.
\newblock \emph{\bibinfo{journal}{Phys. Rev. B}} \textbf{\bibinfo{volume}{75}},
  \bibinfo{pages}{121306};
\newblock \emph{DOI:10.1103/PhysRevB.75.121306} (\bibinfo{year}{2007}).

\bibitem{Ando2013}
\bibinfo{author}{Ando, Y.}
\newblock \bibinfo{title}{{Topological Insulator Materials}}.
\newblock \emph{\bibinfo{journal}{J. Phys. Soc. Jpn.}}
  \textbf{\bibinfo{volume}{82}}, \bibinfo{pages}{102001};
\newblock \emph{DOI:10.7566/JPSJ.82.102001}  (\bibinfo{year}{2013}).

\bibitem{Fu2008}
\bibinfo{author}{Fu, L.} \& \bibinfo{author}{Kane, C.}
\newblock \bibinfo{title}{{Superconducting Proximity Effect and Majorana
  Fermions at the Surface of a Topological Insulator}}.
\newblock \emph{\bibinfo{journal}{Phys. Rev. Lett.}}
  \textbf{\bibinfo{volume}{100}}, \bibinfo{pages}{096407};
\newblock \emph{DOI:10.1103/PhysRevLett.100.096407} (\bibinfo{year}{2008}).

\bibitem{Tanaka2009}
\bibinfo{author}{Tanaka, Y.}, \bibinfo{author}{Yokoyama, T.},
  \bibinfo{author}{Balatsky, A.} \& \bibinfo{author}{Nagaosa, N.}
\newblock \bibinfo{title}{{Theory of topological spin current in
  noncentrosymmetric superconductors}}.
\newblock \emph{\bibinfo{journal}{Phys. Rev. B}} \textbf{\bibinfo{volume}{79}},
  \bibinfo{pages}{060505}; 
\newblock \emph{DOI:10.1103/PhysRevB.79.060505} (\bibinfo{year}{2009}).

\bibitem{Akhmerov2009}
\bibinfo{author}{Akhmerov, A.}, \bibinfo{author}{Nilsson, J.} \&
  \bibinfo{author}{Beenakker, C.}
\newblock \bibinfo{title}{{Electrically Detected Interferometry of Majorana
  Fermions in a Topological Insulator}}.
\newblock \emph{\bibinfo{journal}{Phys. Rev. Lett.}}
  \textbf{\bibinfo{volume}{102}}, \bibinfo{pages}{216404};
\newblock \emph{DOI:10.1103/PhysRevLett.102.216404} (\bibinfo{year}{2009}).

\bibitem{Hasan2010a}
\bibinfo{author}{Hasan, M.~Z.} \& \bibinfo{author}{Kane, C.~L.}
\newblock \bibinfo{title}{{Colloquium: Topological insulators}}.
\newblock \emph{\bibinfo{journal}{Rev. Mod. Phys.}}
  \textbf{\bibinfo{volume}{82}}, \bibinfo{pages}{3045--67};
\newblock \emph{DOI:10.1103/RevModPhys.82.3045} (\bibinfo{year}{2010}).

\bibitem{Qi2011a}
\bibinfo{author}{Qi, X.-L.} \& \bibinfo{author}{Zhang, S.-C.}
\newblock \bibinfo{title}{{Topological insulators and superconductors}}.
\newblock \emph{\bibinfo{journal}{Rev. Mod. Phys.}}
  \textbf{\bibinfo{volume}{83}}, \bibinfo{pages}{1057--110};
\newblock \emph{DOI:10.1103/RevModPhys.83.1057} (\bibinfo{year}{2011}).

\bibitem{Fu2010}
\bibinfo{author}{Fu, L.} \& \bibinfo{author}{Berg, E.}
\newblock \bibinfo{title}{{Odd-Parity Topological Superconductors: Theory and
  Application to Cu$_x$Bi$_{2}$Se$_{3}$}}.
\newblock \emph{\bibinfo{journal}{Phys. Rev. Lett.}}
  \textbf{\bibinfo{volume}{105}}, \bibinfo{pages}{097001};
\newblock \emph{DOI:10.1103/PhysRevLett.105.097001} (\bibinfo{year}{2010}).

\bibitem{Sasaki2011}
\bibinfo{author}{Sasaki, S.} \emph{et~al.}
\newblock \bibinfo{title}{{Topological Superconductivity in
  Cu$_x$Bi$_{2}$Se$_{3}$}}.
\newblock \emph{\bibinfo{journal}{Phys. Rev. Lett.}}
  \textbf{\bibinfo{volume}{107}}, \bibinfo{pages}{217001};
\newblock \emph{DOI:10.1103/PhysRevLett.107.217001} (\bibinfo{year}{2011}).

\bibitem{Kriener2011b}
\bibinfo{author}{Kriener, M.}, \bibinfo{author}{Segawa, K.},
  \bibinfo{author}{Ren, Z.}, \bibinfo{author}{Sasaki, S.} \&
  \bibinfo{author}{Ando, Y.}
\newblock \bibinfo{title}{{Bulk Superconducting Phase with a Full Energy Gap in
  the Doped Topological Insulator Cu$_x$Bi$_2$Se$_3$}}.
\newblock \emph{\bibinfo{journal}{Phys. Rev. Lett.}}
  \textbf{\bibinfo{volume}{106}}, \bibinfo{pages}{127004};
\newblock \emph{DOI:10.1103/PhysRevLett.106.127004} (\bibinfo{year}{2011}).

\bibitem{Hsieh2012}
\bibinfo{author}{Hsieh, T.~H.} \& \bibinfo{author}{Fu, L.}
\newblock \bibinfo{title}{{Majorana Fermions and Exotic Surface Andreev Bound
  States in Topological Superconductors: Application to
  Cu$_x$Bi$_{2}$Se$_{3}$}}.
\newblock \emph{\bibinfo{journal}{Phys. Rev. Lett.}}
  \textbf{\bibinfo{volume}{108}}, \bibinfo{pages}{107005};
\newblock \emph{DOI:10.1103/PhysRevLett.108.107005}
(\bibinfo{year}{2012}).

\bibitem{Goll2002}
\bibinfo{author}{Goll, G.} \emph{et~al.}
\newblock \bibinfo{title}{{Temperature-dependent Fermi surface in CeBiPt}}.
\newblock \emph{\bibinfo{journal}{Europhys. Lett.}}
  \textbf{\bibinfo{volume}{57}}, \bibinfo{pages}{233--9};
\newblock \emph{DOI:10.1209/epl/i2002-00566-9}
(\bibinfo{year}{2002}).

\bibitem{Tafti2013}
\bibinfo{author}{Tafti, F.~F.} \emph{et~al.}
\newblock \bibinfo{title}{{Superconductivity in the noncentrosymmetric
  half-Heusler compound LuPtBi: A candidate for topological
  superconductivity}}.
\newblock \emph{\bibinfo{journal}{Phys. Rev. B}} \textbf{\bibinfo{volume}{87}},
  \bibinfo{pages}{184504};
\newblock \emph{DOI:10.1103/PhysRevB.87.184504}
(\bibinfo{year}{2013}).

\bibitem{Butch2011a}
\bibinfo{author}{Butch, N.~P.}, \bibinfo{author}{Syers, P.},
  \bibinfo{author}{Kirshenbaum, K.}, \bibinfo{author}{Hope, A.~P.} \&
  \bibinfo{author}{Paglione, J.}
\newblock \bibinfo{title}{{Superconductivity in the topological semimetal
  YPtBi}}.
\newblock \emph{\bibinfo{journal}{Phys. Rev. B}} \textbf{\bibinfo{volume}{84}},
  \bibinfo{pages}{220504};
\newblock \emph{DOI:10.1103/PhysRevB.84.220504} 
(\bibinfo{year}{2011}).

\bibitem{Bay2012}
\bibinfo{author}{Bay, T.~V.}, \bibinfo{author}{Naka, T.},
  \bibinfo{author}{Huang, Y.~K.} \& \bibinfo{author}{de~Visser, A.}
\newblock \bibinfo{title}{{Superconductivity in noncentrosymmetric YPtBi under
  pressure}}.
\newblock \emph{\bibinfo{journal}{Phys. Rev. B}} \textbf{\bibinfo{volume}{86}},
  \bibinfo{pages}{064515};
\newblock \emph{DOI:10.1103/PhysRevB.86.064515} 
(\bibinfo{year}{2012}).

\bibitem{Bay2014}
\bibinfo{author}{Bay, T.} \emph{et~al.}
\newblock \bibinfo{title}{{Low field magnetic response of the
  non-centrosymmetric superconductor YPtBi}}.
\newblock \emph{\bibinfo{journal}{Solid State Commun.}}
  \textbf{\bibinfo{volume}{183}}, \bibinfo{pages}{13--7};
\newblock \emph{DOI:10.1016/j.ssc.2013.12.010} 
(\bibinfo{year}{2014}).

\bibitem{Canfield1991}
\bibinfo{author}{Canfield, P.~C.} \emph{et~al.}
\newblock \bibinfo{title}{{Magnetism and heavy fermion-like behavior in the
  RBiPt series}}.
\newblock \emph{\bibinfo{journal}{J. Appl. Phys.}}
  \textbf{\bibinfo{volume}{70}}, \bibinfo{pages}{5800};
\newblock \emph{DOI:10.1063/1.350141} 
(\bibinfo{year}{1991}).

\bibitem{Riedemann1996}
\bibinfo{author}{Riedemann, T.~M.}
\newblock \emph{\bibinfo{title}{{Heat capacities, magnetic properties, and
  resistivities of ternary RPdBi alloys where R = La, Nd, Gd, Dy, Er, and
  Lu}}}.
\newblock MSc thesis, \bibinfo{school}{Iowa State University},
  \bibinfo{address}{Ames}
\newblock \emph{DOI:10.2172/251374} 
(\bibinfo{year}{1996}).

\bibitem{Gofryk2005}
\bibinfo{author}{Gofryk, K.}, \bibinfo{author}{Kaczorowski, D.},
  \bibinfo{author}{Plackowski, T.}, \bibinfo{author}{Leithe-Jasper, A.} \&
  \bibinfo{author}{Grin, Y.}
\newblock \bibinfo{title}{{Magnetic and transport properties of the
  rare-earth-based Heusler phases RPdZ and RPd$_2$Z (Z=Sb,Bi)}}.
\newblock \emph{\bibinfo{journal}{Phys. Rev. B}} \textbf{\bibinfo{volume}{72}},
  \bibinfo{pages}{094409};
\newblock \emph{DOI:10.1103/PhysRevB.72.094409} 
(\bibinfo{year}{2005}).

\bibitem{Kaczorowski2005}
\bibinfo{author}{Kaczorowski, D.}, \bibinfo{author}{Gofryk, K.},
  \bibinfo{author}{Plackowski, T.}, \bibinfo{author}{Leithe-Jasper, A.} \&
  \bibinfo{author}{Grin, Y.}
\newblock \bibinfo{title}{{Unusual features of erbium-based Heusler phases}}.
\newblock \emph{\bibinfo{journal}{J. Magn. Magn. Mater.}}
  \textbf{\bibinfo{volume}{290-291}}, \bibinfo{pages}{573--9};
\newblock \emph{DOI:10.1016/j.jmmm.2004.11.538} 
(\bibinfo{year}{2005}).

\bibitem{Gofryk2011}
\bibinfo{author}{Gofryk, K.}, \bibinfo{author}{Kaczorowski, D.},
  \bibinfo{author}{Plackowski, T.}, \bibinfo{author}{Leithe-Jasper, A.} \&
  \bibinfo{author}{Grin, Y.}
\newblock \bibinfo{title}{{Magnetic and transport properties of
  rare-earth-based half-Heusler phases RPdBi: Prospective systems for
  topological quantum phenomena}}.
\newblock \emph{\bibinfo{journal}{Phys. Rev. B}} \textbf{\bibinfo{volume}{84}},
  \bibinfo{pages}{035208}; 
\newblock \emph{DOI:10.1103/PhysRevB.84.035208}
(\bibinfo{year}{2011}).

\bibitem{Wang2013}
\bibinfo{author}{Wang, W.} \emph{et~al.}
\newblock \bibinfo{title}{{Large Linear Magnetoresistance and Shubnikov-de Hass
  Oscillations in Single Crystals of YPdBi Heusler Topological Insulators.}}
\newblock \emph{\bibinfo{journal}{Sci. Rep.}} \textbf{\bibinfo{volume}{3}},
  \bibinfo{pages}{2181};
\newblock \emph{DOI:10.1038/srep02181} 
(\bibinfo{year}{2013}).

\bibitem{Goraus2013}
\bibinfo{author}{Goraus, J.}, \bibinfo{author}{\'{S}lebarski, A.} \&
  \bibinfo{author}{Fija{\l}kowski, M.}
\newblock \bibinfo{title}{{Experimental and theoretical study of CePdBi.}}
\newblock \emph{\bibinfo{journal}{J. Phys. Condens. Matter}}
  \textbf{\bibinfo{volume}{25}}, \bibinfo{pages}{176002};
\newblock \emph{DOI:10.1088/0953-8984/25/17/176002} 
(\bibinfo{year}{2013}).

\bibitem{Pan2013a}
\bibinfo{author}{Pan, Y.} \emph{et~al.}
\newblock \bibinfo{title}{{Superconductivity and magnetic order in the
  noncentrosymmetric half-Heusler compound ErPdBi}}.
\newblock \emph{\bibinfo{journal}{Europhys. Lett.}}
  \textbf{\bibinfo{volume}{104}}, \bibinfo{pages}{27001};
\newblock \\\emph{DOI:10.1209/0295-5075/104/27001} 
(\bibinfo{year}{2013}).

\bibitem{Pavlosiuk2015}
\bibinfo{author}{Pavlosiuk, O.}, \bibinfo{author}{Filar, K.},
  \bibinfo{author}{Wi\'{s}niewski, P.} \& \bibinfo{author}{Kaczorowski, D.}
\newblock \bibinfo{title}{{Magnetic order and SdH effect in half-Heusler phase
  ErPdBi}}.
\newblock \emph{\bibinfo{journal}{Acta Phys. Polon. A}} \bibinfo{pages}{in
  print} (\bibinfo{year}{2015}).

\bibitem{Xu2014a}
\bibinfo{author}{Xu, G.} \emph{et~al.}
\newblock \bibinfo{title}{{Weak Antilocalization Effect and Noncentrosymmetric
  Superconductivity in a Topologically Nontrivial Semimetal LuPdBi.}}
\newblock \emph{\bibinfo{journal}{Sci. Rep.}} \textbf{\bibinfo{volume}{4}},
  \bibinfo{pages}{5709};
\newblock \emph{DOI:10.1038/srep05709} 
(\bibinfo{year}{2014}).

\bibitem{Gofryk2007}
\bibinfo{author}{Gofryk, K.} \emph{et~al.}
\newblock \bibinfo{title}{{Magnetic, transport, and thermal properties of the
  half-Heusler compounds ErPdSb and YPdSb}}.
\newblock \emph{\bibinfo{journal}{Phys. Rev. B}} \textbf{\bibinfo{volume}{75}},
  \bibinfo{pages}{224426};
\newblock \emph{DOI:10.1103/PhysRevB.75.224426} 
(\bibinfo{year}{2007}).

\bibitem{Freudenberger1998}
\bibinfo{author}{Freudenberger, J.} \emph{et~al.}
\newblock \bibinfo{title}{{Superconductivity and disorder in
  Y$_x$Lu$_{1-x}$Ni$_2$B$_2$C}}.
\newblock \emph{\bibinfo{journal}{Physica C}} \textbf{\bibinfo{volume}{306}},
  \bibinfo{pages}{1--6};
\newblock \emph{DOI:10.1016/S0921-4534(98)00354-2}
(\bibinfo{year}{1998}).

\bibitem{Mueller2001}
\bibinfo{author}{M\"uller, K.~H.} \emph{et~al.}
\newblock \bibinfo{title}{{The upper critical field in superconducting
  MgB$_2$}}.
\newblock \emph{\bibinfo{journal}{J. Alloys Compd.}}
  \textbf{\bibinfo{volume}{322}}, \bibinfo{pages}{L10--13};
\newblock \emph{DOI:10.1016/S0925-8388(01)01197-5} 
(\bibinfo{year}{2001}).

\bibitem{Shulga1998}
\bibinfo{author}{Shulga, S.} \emph{et~al.}
\newblock \bibinfo{title}{{Upper Critical Field Peculiarities of
  Superconducting YNi$_2$B$_2$C and LuNi$_2$B$_2$C}}.
\newblock \emph{\bibinfo{journal}{Phys. Rev. Lett.}}
  \textbf{\bibinfo{volume}{80}}, \bibinfo{pages}{1730--3};
\newblock \emph{DOI:10.1103/PhysRevLett.80.1730} 
(\bibinfo{year}{1998}).

\bibitem{Werthamer1966}
\bibinfo{author}{Werthamer, N.~R.}, \bibinfo{author}{Helfand, E.} \&
  \bibinfo{author}{Hohenberg, P.~C.}
\newblock \bibinfo{title}{{Temperature and Purity Dependence of the
  Superconducting Critical Field, H$_{c2}$. III. Electron Spin and Spin-Orbit
  Effects}}.
\newblock \emph{\bibinfo{journal}{Phys. Rev.}} \textbf{\bibinfo{volume}{147}},
  \bibinfo{pages}{295--302};
\newblock \emph{DOI:10.1103/PhysRev.147.295} 
(\bibinfo{year}{1966}).

\bibitem{Mun2013}
\bibinfo{author}{Mun, E.} \emph{et~al.}
\newblock \bibinfo{title}{{Magnetic-field-tuned quantum criticality of the
  heavy-fermion system YbPtBi}}.
\newblock \emph{\bibinfo{journal}{Phys. Rev. B}} \textbf{\bibinfo{volume}{87}},
  \bibinfo{pages}{075120};
\newblock \emph{DOI:10.1103/PhysRevB.87.075120} 
(\bibinfo{year}{2013}).

\bibitem{Pagliuso2013}
\bibinfo{author}{Pagliuso, P.} \emph{et~al.}
\newblock \bibinfo{title}{{Low temperature specific heat of YBiPt.}}
\newblock \emph{APS March Meeting 2013, Baltimore}, \bibinfo{pages}{Abstract ID: BAPS.2013.MAR.B13.13} 
(\bibinfo{year}{2013}).
\newblock
  \urlprefix\url{http://meeting.aps.org/Meeting/MAR13/Session/B13.13} 
Date of access:15/01/2015.

\bibitem{Pagliuso1999}
\bibinfo{author}{Pagliuso, P.} \emph{et~al.}
\newblock \bibinfo{title}{{Crystal-field study in rare-earth-doped
  semiconducting YBiPt}}.
\newblock \emph{\bibinfo{journal}{Phys. Rev. B}} \textbf{\bibinfo{volume}{60}},
  \bibinfo{pages}{4176--80};
\newblock \emph{DOI:10.1103/PhysRevB.60.4176} 
(\bibinfo{year}{1999}).

\bibitem{Abrikosov1988}
\bibinfo{author}{Abrikosov, A.}
\newblock \emph{\bibinfo{title}{{Fundamentals of the theory of metals}}}
  (\bibinfo{publisher}{North-Holland}, \bibinfo{address}{Amsterdam},
  \bibinfo{year}{1988}).

\bibitem{Petrovic2003}
\bibinfo{author}{Petrovic, C.} \emph{et~al.}
\newblock \bibinfo{title}{{Anisotropy and large magnetoresistance in the
  narrow-gap semiconductor FeSb$_2$}}.
\newblock \emph{\bibinfo{journal}{Phys. Rev. B}} \textbf{\bibinfo{volume}{67}},
  \bibinfo{pages}{155205};
\newblock \emph{DOI:10.1103/PhysRevB.67.155205} 
(\bibinfo{year}{2003}).

\bibitem{Hu2008}
\bibinfo{author}{Hu, J.} \& \bibinfo{author}{Rosenbaum, T.~F.}
\newblock \bibinfo{title}{{Classical and quantum routes to linear
  magnetoresistance.}}
\newblock \emph{\bibinfo{journal}{Nat. Mater.}} \textbf{\bibinfo{volume}{7}},
  \bibinfo{pages}{697--700};
\newblock \emph{DOI:10.1038/nmat2259} 
(\bibinfo{year}{2008}).

\bibitem{Xu1997}
\bibinfo{author}{Xu, R.} \emph{et~al.}
\newblock \bibinfo{title}{{Large magnetoresistance in non-magnetic silver
  chalcogenides}}.
\newblock \emph{\bibinfo{journal}{Nature}} \textbf{\bibinfo{volume}{390}},
  \bibinfo{pages}{57--60};
\newblock \emph{DOI:10.1038/36306} 
(\bibinfo{year}{1997}).

\bibitem{Friedman2010}
\bibinfo{author}{Friedman, A.~L.} \emph{et~al.}
\newblock \bibinfo{title}{{Quantum linear magnetoresistance in multilayer
  epitaxial graphene.}}
\newblock \emph{\bibinfo{journal}{Nano Lett.}} \textbf{\bibinfo{volume}{10}},
  \bibinfo{pages}{3962--5};
\newblock \emph{DOI:10.1021/nl101797d} 
(\bibinfo{year}{2010}).

\bibitem{Wang2012f}
\bibinfo{author}{Wang, X.}, \bibinfo{author}{Du, Y.}, \bibinfo{author}{Dou, S.}
  \& \bibinfo{author}{Zhang, C.}
\newblock \bibinfo{title}{{Room Temperature Giant and Linear Magnetoresistance
  in Topological Insulator Bi$_{2}$Te$_{3}$ Nanosheets}}.
\newblock \emph{\bibinfo{journal}{Phys. Rev. Lett.}}
  \textbf{\bibinfo{volume}{108}}, \bibinfo{pages}{266806};
\newblock \emph{DOI:10.1103/PhysRevLett.108.266806} 
(\bibinfo{year}{2012}).

\bibitem{Shekhar2012b}
\bibinfo{author}{Shekhar, C.} \emph{et~al.}
\newblock \bibinfo{title}{{Electronic structure and linear magnetoresistance of
  the gapless topological insulator PtLuSb}}
\newblock \emph{\bibinfo{journal}{Appl. Phys. Lett.}}
  \textbf{\bibinfo{volume}{100}}, \bibinfo{pages}{252109};
\newblock \emph{DOI:10.1063/1.4730387} 
(\bibinfo{year}{2012}).

\bibitem{Zhang2012}
\bibinfo{author}{Zhang, S.~X.} \emph{et~al.}
\newblock \bibinfo{title}{{Magneto-resistance up to 60 Tesla in topological
  insulator Bi$_2$Te$_3$ thin films}}
\newblock \emph{\bibinfo{journal}{Appl. Phys. Lett.}}
  \textbf{\bibinfo{volume}{101}}, \bibinfo{pages}{202403}; 
\newblock \emph{DOI:10.1063/1.4766739} 
(\bibinfo{year}{2012}).

\bibitem{He2014}
\bibinfo{author}{He, L.~P.} \emph{et~al.}
\newblock \bibinfo{title}{{Quantum transport evidence for a three-dimensional
Dirac semimetal phase in Cd$_3$As$_2$}} 
\newblock \emph{\bibinfo{journal}{Phys. Rev. Lett.}}
  \textbf{\bibinfo{volume}{113}}, \bibinfo{pages}{246402}; 
\newblock \emph{DOI:10.1103/PhysRevLett.113.246402} 
(\bibinfo{year}{2014}). 

\bibitem{Novak2014}
\bibinfo{author}{Novak, M.}, \bibinfo{author}{Sasaki, S.},
  \bibinfo{author}{Segawa, K.} \& \bibinfo{author}{Ando, Y.}
\newblock \bibinfo{title}{{Large linear magnetoresistance in the Dirac
  semimetal TlBiSSe}} 
\newblock \eprint[arXiv]{preprint arXiv:1408.2183} (\bibinfo{year}{2014}).
\newblock \urlprefix\url{http://lanl.arxiv.org/abs/1408.2183} Date of access:15/01/2015.

\bibitem{Parish2003}
\bibinfo{author}{Parish, M.~M.} \& \bibinfo{author}{Littlewood, P.~B.}
\newblock \bibinfo{title}{{Non-saturating magnetoresistance in heavily
  disordered semiconductors}}.
\newblock \emph{\bibinfo{journal}{Nature}} \textbf{\bibinfo{volume}{426}},
  \bibinfo{pages}{1--4}; 
\newblock \emph{DOI:10.1038/nature02073} 
(\bibinfo{year}{2003}).

\bibitem{Abrikosov1998a}
\bibinfo{author}{Abrikosov, A.}
\newblock \bibinfo{title}{{Quantum magnetoresistance}}.
\newblock \emph{\bibinfo{journal}{Phys. Rev. B}} \textbf{\bibinfo{volume}{58}},
  \bibinfo{pages}{2788--94};
\newblock \emph{DOI:10.1103/PhysRevB.58.2788} 
(\bibinfo{year}{1998}).

\bibitem{Hikami1980}
\bibinfo{author}{Hikami, S.}, \bibinfo{author}{Larkin, A.~I.} \&
  \bibinfo{author}{Nagaoka, Y.}
\newblock \bibinfo{title}{{Spin-Orbit Interaction and Magnetoresistance in the
  Two Dimensional Random System}}.
\newblock \emph{\bibinfo{journal}{Prog. Theor. Phys.}}
  \textbf{\bibinfo{volume}{63}}, \bibinfo{pages}{707--10};
\newblock \emph{DOI:10.1143/PTP.63.707} 
(\bibinfo{year}{1980}).

\bibitem{Shoenberg1984}
\bibinfo{author}{Shoenberg, D.}
\newblock \emph{\bibinfo{title}{{Magnetic Oscillations in Metals}}}
  (\bibinfo{publisher}{Cambridge University Press},
  \bibinfo{address}{Cambridge}, \bibinfo{year}{1984}).

\bibitem{Analytis2010}
\bibinfo{author}{Analytis, J.~G.} \emph{et~al.}
\newblock \bibinfo{title}{{Two-dimensional surface state in the quantum limit
  of a topological insulator}}.
\newblock \emph{\bibinfo{journal}{Nature Phys.}} \textbf{\bibinfo{volume}{6}},
  \bibinfo{pages}{960--4};
\newblock \emph{DOI:10.1038/nphys1861} 
(\bibinfo{year}{2010}).

\bibitem{Chen2011}
\bibinfo{author}{Chen, J.} \emph{et~al.}
\newblock \bibinfo{title}{{Tunable surface conductivity in Bi$_{2}$Se$_{3}$
  revealed in diffusive electron transport}}.
\newblock \emph{\bibinfo{journal}{Phys. Rev. B}} \textbf{\bibinfo{volume}{83}},
  \bibinfo{pages}{241304};
\newblock \emph{DOI:10.1103/PhysRevB.83.241304} 
(\bibinfo{year}{2011}).

\bibitem{Taskin2011b}
\bibinfo{author}{Taskin, A.~A.} \& \bibinfo{author}{Ando, Y.}
\newblock \bibinfo{title}{{Berry phase of nonideal Dirac fermions in
  topological insulators}}.
\newblock \emph{\bibinfo{journal}{Phys. Rev. B}} \textbf{\bibinfo{volume}{84}},
  \bibinfo{pages}{035301}; 
\newblock \emph{DOI:10.1103/PhysRevB.84.035301} 
(\bibinfo{year}{2011}).

\bibitem{Rodriguez-Carvajal1993}
\bibinfo{author}{Rodr\'{\i}guez-Carvajal, J.}
\newblock \bibinfo{title}{{Recent advances in magnetic structure determination
  by neutron powder diffraction}}.
\newblock \emph{\bibinfo{journal}{Physica B Condens. Matter}}
  \textbf{\bibinfo{volume}{192}}, \bibinfo{pages}{55--69}; 
\newblock \\\emph{DOI:10.1016/0921-4526(93)90108-I} 
(\bibinfo{year}{1993}).

\bibitem{Sheldrick2008}
\bibinfo{author}{Sheldrick, G.~M.}
\newblock \bibinfo{title}{{A short history of SHELX.}}
\newblock \emph{\bibinfo{journal}{Acta Crystallogr. A.}}
  \textbf{\bibinfo{volume}{64}}, \bibinfo{pages}{112--22};
\newblock \emph{DOI:10.1107/S0108767307043930} 
(\bibinfo{year}{2008}).
\end{thebibliography}

\begin{thebibliography}{1}
\expandafter\ifx\csname url\endcsname\relax
  \def\url#1{\texttt{#1}}\fi
\expandafter\ifx\csname urlprefix\endcsname\relax\def\urlprefix{URL }\fi
\providecommand{\bibinfo}[2]{#2}
\providecommand{\eprint}[2][]{\url{#2}}

\bibitem{Haase2002}
\bibinfo{author}{Haase, M.~G.}, \bibinfo{author}{Schmidt, T.},
  \bibinfo{author}{Richter, C.~G.}, \bibinfo{author}{Block, H.} \&
  \bibinfo{author}{Jeitschko, W.}
\newblock \bibinfo{title}{{Equiatomic Rare Earth (Ln) Transition Metal
  Antimonides LnTSb (T=Rh, lr) and Bismuthides LnTBi (T=Rh, Ni, Pd, Pt)}}.
\newblock \emph{\bibinfo{journal}{J. Solid State Chem.}}
  \textbf{\bibinfo{volume}{168}}, \bibinfo{pages}{18--27};
\newblock \emph{DOI:10.1006/jssc.2002.9670} (\bibinfo{year}{2002}).

\bibitem{Shan2013}
\bibinfo{author}{Shan, R.} \emph{et~al.}
\newblock \bibinfo{title}{{Electronic and crystalline structures of zero
  band-gap LuPdBi thin films grown epitaxially on MgO(100)}}.
\newblock \emph{\bibinfo{journal}{Appl. Phys. Lett.}}
  \textbf{\bibinfo{volume}{102}}, \bibinfo{pages}{172401};
\newblock \emph{DOI:10.1063/1.4802795} (\bibinfo{year}{2013}).

\bibitem{Xu2014a}
\bibinfo{author}{Xu, G.} \emph{et~al.}
\newblock \bibinfo{title}{{Weak Antilocalization Effect and Noncentrosymmetric
  Superconductivity in a Topologically Nontrivial Semimetal LuPdBi.}}
\newblock \emph{\bibinfo{journal}{Sci. Rep.}} \textbf{\bibinfo{volume}{4}},
  \bibinfo{pages}{5709};
\newblock \emph{DOI:10.1038/srep05709} (\bibinfo{year}{2014}).

\bibitem{Nowak2014}
\bibinfo{author}{Nowak, B. \& Kaczorowski, D.} 
\newblock \bibinfo{title}{{NMR as a Probe of Band Inversion in Topologically 
 Nontrivial Half-Heusler Compounds.}}
 \newblock \emph{\bibinfo{journal}{J. Phys. Chem. C}} \textbf{\bibinfo{volume}{118}},
  \bibinfo{pages}{18021};
\newblock \emph{DOI:10.1021/jp505320w} (\bibinfo{year}{2014}).

\bibitem{Hou2015}
\bibinfo{author}{Hou, Z.}  \emph{et~al.}
\newblock \bibinfo{title}{{Transition from semiconducting to metallic-like 
conducting and weak antilocalization effect in single crystals of LuPtSb.}},
\newblock \eprint[arXiv]{preprint arXiv:1501.06714} (\bibinfo{year}{2015}).
\newblock \urlprefix\url{http://arxiv.org/abs/1501.06714} Date of access: 15/02/2015.

\bibitem{Shekhar2015}
\bibinfo{author}{Shekhar, C., Kampert, E., F\"{o}rster, T., Nayak, A.K., Nicklas, M. \& Felser, C.}
\newblock \bibinfo{title}{{Large linear magnetoresistance and weak anti-localizatio n in Y(Lu)PtBi topological insulators.}}
\newblock \eprint[arXiv]{preprint arXiv:1502.00604} (\bibinfo{year}{2015}).
\newblock \urlprefix\url{http://arxiv.org/abs/1502.00604} Date of access: 15/02/2015.

\bibitem{Mun2013}
\bibinfo{author}{Mun, E.} \emph{et~al.}
\newblock \bibinfo{title}{{Magnetic-field-tuned quantum criticality of the
  heavy-fermion system YbPtBi}}.
\newblock \emph{\bibinfo{journal}{Phys. Rev. B}} \textbf{\bibinfo{volume}{87}},
  \bibinfo{pages}{075120}; 
\newblock \emph{DOI:10.1103/PhysRevB.87.075120} (\bibinfo{year}{2013}).

\bibitem{Pagliuso2013}
\bibinfo{author}{Pagliuso, P.} \emph{et~al.}
\newblock \bibinfo{title}{{Low temperature specific heat of YBiPt.}}
\newblock \emph{APS March Meeting 2013, Baltimore}, \bibinfo{pages}{Abstract ID: BAPS.2013.MAR.B13.13} (\bibinfo{year}{2013}).
\newblock
  \urlprefix\url{http://meeting.aps.org/Meeting/MAR13/Session/B13.13} 
Date of access:15/01/2015.

\bibitem{Pavlosiuk_YPtBi}
\bibinfo{author}{Pavlosiuk, O.}, \bibinfo{author}{Wi\'{s}niewski, P.} \&
  \bibinfo{author}{Kaczorowski, D.}
\newblock \bibinfo{title}{{Superconductivity and Shubnikov-de Haas oscillations 
in the noncentrosymetric half-Heusler compound YPtBi.}}
\newblock \bibinfo{note}{SCES'14 Conference, Grenoble} \bibinfo{pages}{Abstract Mo-101} (\bibinfo{year}{2014}).
\newblock 
 \urlprefix\url{https://www.ill.eu/nc/fr/presse-et-infos/events/sces-2014/abstract-booklet/?cid=45039&did=69737&sechash=f08eb1b1} Date of access:15/01/2015.

\bibitem{Riedemann1996}
\bibinfo{author}{Riedemann, T.~M.}
\newblock \emph{\bibinfo{title}{{Heat capacities, magnetic properties, and
  resistivities of ternary RPdBi alloys where R = La, Nd, Gd, Dy, Er, and
  Lu}}}.
\newblock MSc thesis, \bibinfo{school}{Iowa State University},
  \bibinfo{address}{Ames}
\newblock \emph{DOI:10.2172/251374} 
(\bibinfo{year}{1996}).

\bibitem{Nowak2013}
\bibinfo{author}{Nowak, B.} \& \bibinfo{author}{Kaczorowski, D.}
\newblock \bibinfo{title}{{Nonmetallic behaviour in half-Heusler phases YPdSb,
  YPtSb and LuPtSb}}.
\newblock \emph{\bibinfo{journal}{Intermetallics}}
  \textbf{\bibinfo{volume}{40}}, \bibinfo{pages}{28--35};
\newblock \emph{DOI:10.1016/j.intermet.2013.04.001} (\bibinfo{year}{2013}).

\end{thebibliography}

\end{document}